\documentclass[12pt]{article}
\usepackage{amsmath,amssymb,amsfonts}
\usepackage{iiit_thesis}
\usepackage{times}
\usepackage{graphicx}
\usepackage{setspace}
\usepackage{enumitem}
\usepackage{tabularx}
\usepackage{subfigure}
\usepackage{notoccite}
\usepackage{afterpage}
\usepackage{xcolor}
\usepackage{algorithm}
\usepackage{algpseudocode}
\usepackage{multirow}
\usepackage[hidelinks]{hyperref} 
\usepackage[toc,page]{appendix}
\usepackage{comment}
\usepackage{pdfpages}
\usepackage{siunitx}
\usepackage{authblk}
\usepackage[style=numeric-comp,sorting=none,citestyle=numeric-comp, maxnames=99]{biblatex}
\renewbibmacro{in:}{}
\DeclareFieldFormat[article]{pages}{#1}
\let\oldcite\cite
\renewcommand{\cite}[1]{\textsuperscript{\oldcite{#1}}}
\graphicspath{{images/}{./}} 

\renewcommand{\baselinestretch}{1.5} 
\title{\textbf{High Thermoelectric Performance via Stacking-Controlled Symmetry Breaking in Layered XZnBi (X = Rb, Cs) Zintl Materials}}

\author[1]{Aadil Fayaz Wani}
\author[1]{Nirma Kumari}
\author[1]{SuDong Park}
\author[1,2]{*Byungki Ryu }
\affil[1]{\textit{Energy Conversion Research Center, Electrical Materials Research Division, Korea Electrotechnology Research Institute (KERI), Changwon-si, 51543, Republic of Korea}}
\affil[2]{\textit{Electrical-Functionality Materials Engineering, KERI School, University of Science and Technology (UST), Changwon-si, 51543, Republic of Korea}}
\date{\textit{*Email: byungkiryu@keri.re.kr}}

\begin{document}
\maketitle   

\begin{abstract}
High thermoelectric efficiency requires high Seebeck coefficient, high electrical conductivity, and low thermal conductivity. However, strategies that suppress thermal conductivity often simultaneously degrade electrical conductivity, making effective electrical-thermal decoupling highly challenging. Here, we show that atomic-layer stacking order change in XZnBi (X = Rb, Cs) provides an efficient route to achieve such decoupling. Even though electronic transport coefficients and relaxation times remain largely insensitive to stacking order due to preserved Fermi-surface topology, the lattice thermal conductivity exhibits a strong stacking dependence, with AB stacking significantly suppressing it below 1 Wm$^{-1}$K$^{-1}$ at temperatures above 300 K. The stacking transition from AA to AB breaks structural symmetries. It increases the three-phonon phase space and available scattering channel, substantially suppressing phonon transport by about 50$\%$ in both materials. As a result, the AB-stacked phases yield high ZT values of 1.96 (1.69) in n-type CsZnBi (RbZnBi) at  900 K, which is about 40$\%$ (30$\%$) higher than AA stacking. These findings establish the XZnBi family as promising thermoelectric candidates and highlight stacking-order controlled phonon transport as a robust strategy for advancing thermoelectric material design.

\end{abstract}

\section*{Keywords}
\textit{First-principles calculations; Electron-phonon coupling; Stacking-induced anharmonicity; Electrical-thermal decoupling; Thermoelectric properties}

\section[5pt]{Introduction}
In recent years, the increasing demand for sustainable and clean energy technologies has intensified the search for materials with superior thermoelectric performance. Thermoelectric materials enable direct conversion between heat and electricity and hold promise for waste heat recovery, solid-state cooling, and remote power generation\cite{bell2008cooling}. The efficiency of a thermoelectric material is governed by the dimensionless figure of merit;
\begin{equation}
 ZT =  \frac{S^{2}\sigma T}{k_{tot}} 
\end{equation}
where $S$ is the Seebeck coefficient, $\sigma$ is the electrical conductivity, T is the absolute temperature, and $k$ is the total thermal conductivity, comprising both lattice ($k_{l}$) and electronic ($k_{e}$) contributions\cite{hsu2004cubic, snyder2008complex,ryu2021thermoelectric,ryu2025thermoelectric}. For a thermoelectric material to achieve high performance, a material must simultaneously exhibit a large Seebeck coefficient, high electrical conductivity, and low thermal conductivity, an inherently conflicting set of requirements due to the interdependence of these parameters via the charge carrier concentration\cite{xiao2014decoupling}. The development of thermoelectric materials with high efficiency and stability at elevated temperatures is essential for advancing sustainable energy technologies\cite{singh2024advancements, weidenkaff2017thermoelectricity}. At elevated temperatures, an optimal electronic bandgap is essential to prevent bipolar effect. Minimizing lattice thermal conductivity is also key to achieving a high $ZT$ value, which can be accomplished through methods such as nano-structuring or alloying\cite{song2025reduced, zhao2011toward}. Doping is also an effective strategy and is widely employed in thermoelectric society to tune the thermoelectric properties of materials\cite{zhao2020chemical}. However, finding a suitable dopant is challenging and often requires extensive computational or experimental studies\cite{ayachi2024high,lee2019fine}. Consequently, materials that intrinsically exhibit low lattice thermal conductivity and favorable electronic transport characteristics are particularly attractive. These include complex crystal structures with heavy atoms, low phonon group velocities, or layered structures that strongly scatter heat-carrying phonons\cite{zhao2014ultralow}. This unique combination offers a promising pathway toward high-performance thermoelectric devices with improved thermal stability and efficiency under harsh operating conditions. 

A central challenge in the development of high-efficiency thermoelectrics is the inherent coupling between electrical and thermal transport, which typically forces a compromise between a high power factor ($S^2\sigma$) and low thermal conductivity ($\kappa$). Achieving "electron-crystal, phonon-glass" behavior requires strategies that can selectively disrupt phonon transport without impeding the movement of charge carriers\cite{wu2013origin, zhu2018thermoelectric}. In this context, layered materials offer a unique platform for such decoupling, as their anisotropic bonding allows independent tuning of cross-plane and in-plane interactions. Recent studies on van der Waals and Zintl-phase materials have demonstrated that stacking order and interlayer orientation can serve as powerful degrees of freedom to modulate the vibrational and thermal properties\cite{li2022layered,wang2017thermal}. For instance, in materials like MoS$_2$ and WSe$_2$, transitions between different stacking arrangements have been shown to significantly suppress $\kappa_l$ by altering the symmetry-dependent scattering phase space\cite{mk2023layer,chiritescu2007ultralow}. The stacking order-induced symmetry reduction causes a dramatic reduction in $\kappa_l$ and the values reach close to the amorphous limit (below 1 Wm$^{-1}$K$^{-1}$). Hence, by manipulating these stacking-dependent properties, phonons can be strongly scattered through enhancing anharmonicity, while maintaining the power factor supported by the robust electronic topology of the layers.

Zintl compounds with ABX 1:1:1 stoichiometry frequently crystallizes in structures like ZrBeSi or PbFCl\cite{schoop2018chemical, owens2020synthesis, xu2015two}. They have emerged as a promising class of thermoelectric materials due to their unique combination of complex crystal structures and low lattice thermal conductivity. Originating from a formal valence electron counting scheme, Zintl phases are formed by the electron transfer from electropositive (alkali or alkaline earth) elements to more electronegative post-transition metal or metalloid frameworks\cite{khireddine2021first, kauzlarich2023zintl, wu2025zintl}. This leads to structurally diverse compounds with semiconducting or semimetallic behavior. Their complex bonding nature and heavy constituent elements result in strong phonon scattering and inherently low lattice thermal conductivity, making them highly attractive for thermoelectric applications. Their band structures often feature multiple valleys, band convergence, and nonparabolic dispersion, which are beneficial for enhancing the Seebeck coefficient and the density of states near the Fermi level\cite{zhang2025anharmonic}. Moreover, their weak interlayer bonding and potential for stacking variations introduce additional degrees of freedom to engineer anisotropic transport properties. The large atomic mass and structural complexity further suppress heat transport via phonons, facilitating the realization of phonon-glass electron-crystal (PGEC) behavior\cite{wei2023rattling, liu2019recent, nolas2001phonon}. These attributes make them particularly compelling for high-temperature thermoelectric applications.

RbZnBi and CsZnBi are recently proposed Zintl-type compounds that exhibit intriguing electronic properties favorable for topological  applications\cite{lee2024large}. According to first-principles calculations, these materials can be stabilized in layered hexagonal structures with either AA or AB stacking of ZnBi layers. Importantly, both RbZnBi and CsZnBi possess sizable spin-orbit coupling-induced bulk band gaps. This large band gap not only ensures the robustness of the topological features at elevated temperatures but also mitigates the bipolar effect, a common challenge in thermoelectric materials. The layered nature, combined with low lattice thermal conductivity predicted for similar Zintl phases, suggests that XZnBi could offer high Seebeck coefficients and enhanced thermoelectric $ZT$, especially when tuned via stacking order. These unique characteristics position XZnBi as promising candidates for next-generation thermoelectric materials that leverage topological protection and electronic tunability.

In this study, we perform a comprehensive first-principles investigation of electronic and thermal transport properties of XZnBi, evaluating both AA and AB stacking configurations. By employing Boltzmann transport theory with temperature-dependent scattering rates derived from phonon-phonon and electron-phonon coupling calculations, we elucidate the role of stacking order, doping type, and temperature on thermoelectric transport. Since constant relaxation time approximation (CRTA) highly overestimates the transport properties, we have calculated a temperature-dependent relaxation time using fully first-principles approach. The results provide insight into the interplay between electronic and phononic band structure, scattering mechanism, and stacking-dependent transport properties and offer design principles for optimizing thermoelectric performance in Zintl compounds.

\section{Theoretical method}
Structural optimization and electronic structure calculations are performed using first-principles calculations within the framework of density functional theory (DFT) using Vienna Ab initio Simulation Package (VASP)\cite{kresse1996efficient} and Quantum Espresso (QE) tools\cite{giannozzi2009quantum}. The initial structures are taken from the reference\cite{lee2024large}. Exchange-correlation effects are treated with the generalized gradient approximation (GGA) using the Perdew-Burke-Ernzerhof (PBE) functional\cite{perdew1996generalized}. The plane-wave basis set is expanded with an energy cutoff of 550 eV to ensure convergence. For Brillouin zone integration, $\Gamma$-centered k-point meshe of 12$\times$12$\times$8 and 12$\times$12$\times$4 are employed for P$\overline{6}$m2 and P6$_3$/mmc crystal structures, respectively\cite{pack1977special}. Structural optimization was carried out until the residual atomic forces converged to less than 10$^{-5}$ eV$\AA ^{-1}$, ensuring accurate lattice parameters, internal coordinates and lattice thermal conductivity. Spin-orbit coupling (SOC) effects are also incorporated in both materials.

Transport properties are evaluated using BoltzTrap2 code, an open-source framework designed to solve the electronic Boltzmann transport equations\cite{madsen2018boltztrap2}. The underlying electron-phonon coupling is computed with the help of the Electron-Phonon Coupling using Wannier functions (EPW) code by calculating electron-phonon interactions\cite{lee2023electron} using Quantum Espresso (QE) package. The fully relativistic norm-conserving pseudopotentials are used to include the SOC effects with similar energy cutoff values used for VASP. Electron-phonon scattering rates are obtained through Wannier interpolation of electron-phonon matrix elements, based on electronic and phonon band structure calculations performed using QE framework\cite{sjakste2015wannier}. The EPW code was utilized to interpolate the electron-phonon interaction matrices onto an ultra-dense fine mesh of \(60\times60\times60\) \(\mathbf{k}\)-points and \(40\times40\times40\) \(\mathbf{q}\)-points to ensure the strict convergence of scattering rates. We employ the rigid band approximation to model the shift in the Fermi level with doping concentration, allowing it to vary from near the valence band maximum (VBM) to near the conduction band minimum (CBM). From the calculated lifetimes, transport coefficients are predicted at the relaxation time approximation (RTA) level using:
\begin{equation}	
	\small
	 \sigma_{(\alpha, \beta)} (T,E_F) = \frac {1} {\Omega} \int  \hat{\sigma}_{(\alpha, \beta)}(\varepsilon)  [-\frac{\partial{f_{0}}(T,\varepsilon,E_{F})}{\partial \varepsilon}] d\varepsilon     
\end{equation}
	and
\begin{equation}
	\small		
	S_{(\alpha, \beta)} (T,E_F) = \frac {1} {eT \sigma_{(\alpha, \beta)} (T,E_F)}		
		 \int (E - E_F) \hat{\sigma}_{(\alpha, \beta)}(\varepsilon)[-\frac{\partial{f_{0}}(T,\varepsilon,E_{F})}{\partial \varepsilon}] d\varepsilon   
\end{equation}
		where $\alpha$, $\beta$ are cartesian indices, $f_{0}$ is Fermi-distribution function and $\Omega$ is unit cell volume. The $\kappa_{e}$ and $\sigma$ are coupled by Wiedmann-Franz law given by:
		$\kappa_{e} = L \sigma T $
		where L is the Lorenz number\cite{kim2015characterization}. The central parameters that govern the distribution of carriers across the crystal is transport distribution function:
\begin{equation}	
	\hat{\sigma}_{(\alpha, \beta)} = \frac{e^2}{N} \sum_{i,k} \tau v_{\alpha}(i,k)v_{\beta}(i,k)\delta(\varepsilon - \varepsilon_{ik})
\end{equation}
	where $i$ and $k$ are band index and wave vector, respectively, and $\tau$ is relaxation time of carriers\cite{ziman2001electrons}. The Brillouin zone integration is carried out using a dense mesh of 200 times the k-point mesh used for band structure calculation, to ensure convergence of the transport coefficients.

The lattice thermal conductivity ($k_{l}$) is obtained by solving linearized phonon Boltzmann transport equation, under the assumption that $k_{l}$ is proportional to the phonon lifetimes\cite{guyer1966solution}. These lifetimes are calculated from lattice anharmonicity using first-principles method. The second order force constants are calculated using phonopy package with finite difference method approach\cite{togo2023implementation}. For third-order force constant calculations, phono3py package is used with supercells of 3$\times$3$\times$3 and 3$\times$3$\times$2 for AA and AB stacking, respectively\cite{togo2015distributions}. For high accuracy, atomic interactions upto 3rd nearest neighbors and dense meshes of 31$\times$31$\times$31 are used for calculating $k_{l}$.

\section{Results and Discussions}
\subsection{Crystal Structure and Chemical Bonding}
\begin{figure}[H]
   \centering    
   \includegraphics[width=0.9\textwidth, height = 12cm]{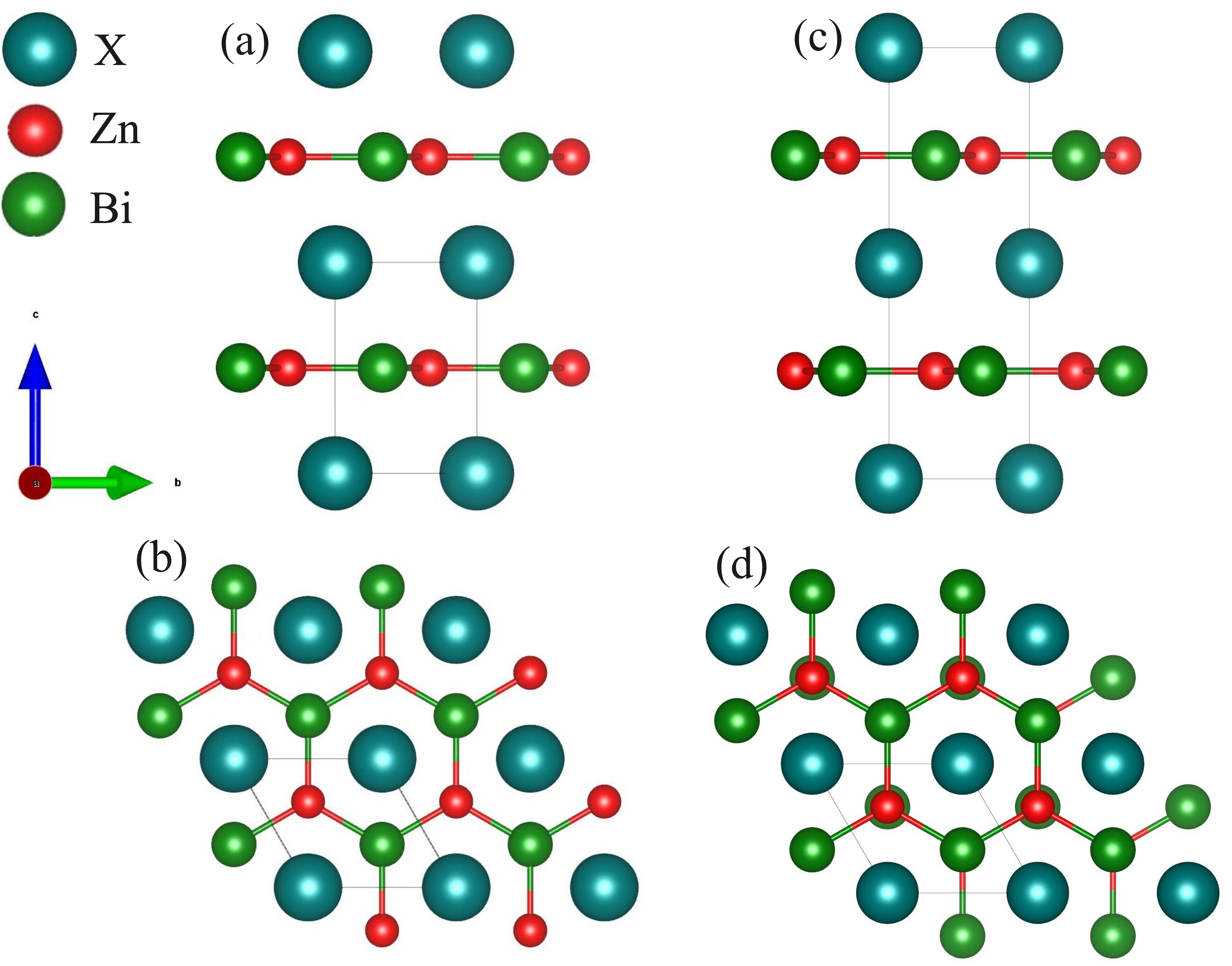}
   \caption{Crystal structure of XZnBi (X = Rb, Cs) in AA stacking (left) and AB stacking (right) as seen along (a, c) x-y direction and (b, d) z-direction. A solid line box shows the unit cell in x-y plane.}
   \label{fig:1.1}
\end{figure}

Two energetically favorable phases with space groups P$\overline{6}$m2 and P6$_3$/mmc have been identified for XZnBi compounds (X=Rb, Cs) as shown in Figure \ref{fig:1.1}. These compounds adopt ZrBeSi structure type, an ordered derivative of the widely known AlB2 structure\cite{matar2019coloring}. Both phases exhibit hexagonal symmetry, where 2D Boron Nitride-like Zn and Bi layers in the xy-plane are sandwiched between alkali metal X atoms positioned at the center of hexagonal rings in adjacent layers. The crystal structure is inherently anisotropic and one may expect direction-dependent transport properties. The saparation between the layers along $z$-direction is around $\sim$ 6 Å (exact values are given in Table \ref{tab:1.0}), with minimal interlayer interaction dominated by Van der Waals (VdW) forces along the z-direction. This weak interlayer bonding may reduce phonon coupling in the $z$-direction, thereby impeding phonon transport across layers. Consequently, these materials are expected to exhibit low lattice thermal conductivity, as will be discussed in forthcoming section. In XZnBi (X=Rb, Cs), the lattice parameter $a$ remains almost constant despite the size difference of the alkali metal, measuring 4.55 $\AA$ for RbZnBi and 4.56 $\AA$ for CsZnBi. Instead, the unit cell expands predominantly along the $c$-axis, corresponding to the interlayer direction, with values of 11.10 $\AA$ for RbZnBi and 11.92 $\AA$ for CsZnBi. This indicates that the alkali metal size primarily influences vertical interlayer spacing while the in-plane ZnBi layer structure remains unaffected due to the energetically unfavorable expansion of the strong Zn–Bi bonds. This lack of variation of lattice parameter $a$ with cation size is a unique feature compared to other systems.

The alkali metals X = Rb/Cs possess a single $s$-electron in their 4$s$/5$s$ orbitals, which they readily lose to attain stability. This electron is transfered to more electronegative Bi atom, forming an ionic bond between X and Bi. This aligns with the Zintl-Klemm concept of zintl compounds, where electropositive elements transfer electrons to an anionic framework\cite{nesper2014zintl}. Meanwhile, Zn forms primarily ionic bonds with Bi, as the filled 3$d$ orbitals of Zn are energetically deep, leading to its 4$s$ electrons to participate in bonding with Bi. The difference in electronegativity between Zn (1.65) and Bi (2.02) reinforces mix ionic-covalent character of these bonds\cite{tantardini2021thermochemical}. The phase formation energy with respect to isolated elemental phases is calculated to evaluate the chemical stability of both compounds. The resulting values are -0.27(-0.28) eV/atom for RbZnBi and -0.31(-0.29) eV/atom for CsZnBi in the AA(AB) stacking, respectively. This demonstrates that the compound phases are thermodynamically favorable.

To validate the above discussion, electron localization function (ELF)\cite{chesnut1999electron} is plotted which provides further insights into the nature of chemical bonding as shown in Figure S1. Regions with low ELF values (blue) correspond to X atoms or interlayer spaces between ZnBi planes, where the alkali metal X donates its single $s$-electron. This results in symmetric, low electron localization around X ions. High ELF values (around 1) around Bi atoms indicate significant electron localization, confirming that Bi captures the $s$-electron from X atom to form ionic bond. The localized electron density around Bi stabilizes the ZnBi layer structure, while ELF values around 0.5 between Zn and Bi suggest a predominantly mix ionic-covalent bond character due to their electronegativity differences. Notably, the ELF around Bi is not spherically symmetric. This asymmetry arises from $s$ valence electrons of Bi, including two non-bonding 6$s$ electrons and two 6$p$ electrons (e.g., p$_{x}$), forming stereochemically active lone pairs. These lone pairs appear as bulged electron densities along the c-axis. The stereochemically active lone pairs in Bi may play a role in increased vibrational anharmonicity, a feature associated with low lattice thermal conductivity\cite{nissimagoudar2020lattice}. This effect is consistent with observations in other well-known thermoelectric materials, such as Bi$_{2}$Te$_{3}$, Sb$_{2}$Te$_{3}$, SnSe, and PbTe, which are renowned for their low thermal conductivity and enhanced thermoelectric performance\cite{zhang2026low,parashchuk2022ultralow,wu2026lone,long2025theoretical,}.

		\begin{table*}[h!]
		\centering
		\caption{Optimized lattice parameters along x (a) and z (c) direction, interlayer spacing (d), and calculated band gap ($E_g$) for XZnBi in AA and AB stacked configurations.}
        \renewcommand{\arraystretch}{1.5}
        \vspace{0.5cm}
		\begin{tabular}{ccccccc}

			\hline
			Material & Stacking & a ($\AA$) & c ($\AA$) & d ($\AA$)  & $E_g$ (eV)  \\

			\hline
			
			\multirow{2}{*}{RbZnBi} & AA & 4.75 & 5.76 & 5.76 & 0.53  \\

			  & AB & 4.76 & 11.41 & 5.70 & 0.45 \\
			
			\multirow{2}{*}{CsZnBi} & AA & 4.75 & 6.16 & 6.16 & 0.50 \\

			& AB & 4.79 & 12.29 & 6.15 & 0.44 \\
			\hline
			\label{tab:1.0}
		\end{tabular}
	\end{table*}
\subsection{Electronic Band Structure}
\begin{figure}[H]
   \centering    
    \includegraphics[width=1\textwidth, height = 14cm]{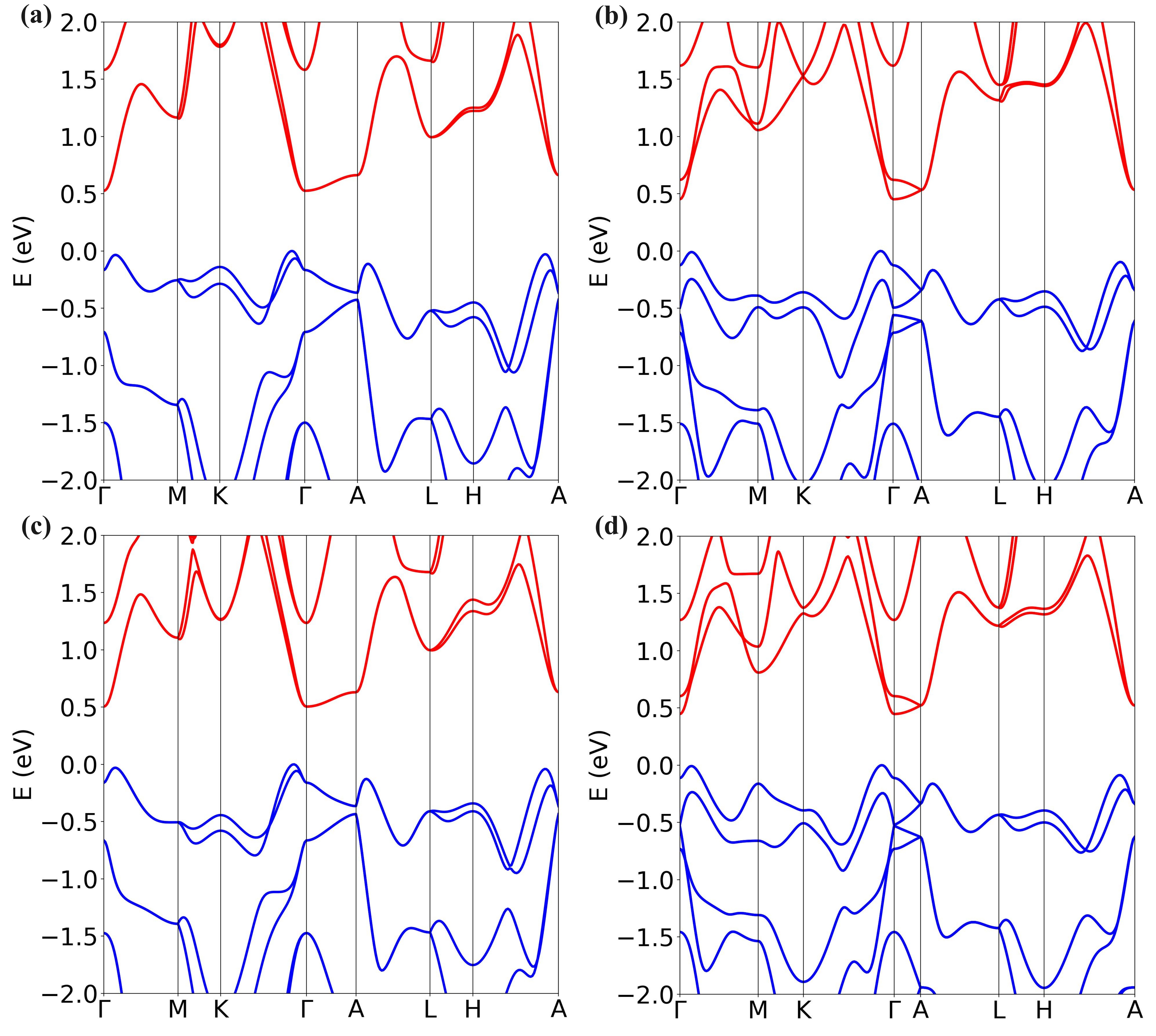}
    \caption{Electronic band structure of (a, b) RbZnBi and (c, d) CsZnBi in AA and AB stacking, respectively, along high-symmetry paths. The $\Gamma$-K and $\Gamma$-M paths are in x-y plane while $\Gamma$-Z is along z-axis.}
    \label{fig:1.2}
\end{figure}

Figure \ref{fig:1.2} presents the electronic band structure of XZnBi compounds in both AA and AB stacking configurations, calculated using PBE functionals with SOC effects. Both compounds exhibit semiconducting behavior with moderate band gaps, as listed in Table \ref{tab:1.0}. The values are in good agreement with previous theoretical studies. 

Let us first analyze the electronic structure of AA-stacked phases based on their orbital bonding. Since the substitution of Rb with Cs results in a subtle change in crystal structure, the electronic band structures of the two compounds are nearly identical, except for a slight decrease in magnitude of bandgap from 0.53 eV (RbZnBi) to 0.50 eV (CsZnBi). This reduction arises from the larger atomic size and increased orbital overlap in CsZnBi compared to RbZnBi, which brings the band edges closer together and narrows the bandgap. The $\Gamma$-K and $\Gamma$-A paths correspond to directions in the x-y plane and along the out-of-plane z-direction, respectively. Stronger in-plane interactions between Zn and Bi atoms lead to highly dispersive bands along the $\Gamma$-K direction, while weaker interlayer interactions, including those involving X atoms along the z-direction, result in flat, low-dispersion bands along the $\Gamma$-A direction as shown in Figure \ref{fig:1.2}. This anisotropic band structure behavior is characteristic of layered materials.

Coming to the AB stacking, the broken crystal symmetry along z-axis affects the interlayer bonding. The result is that the crystal lattice of RbZnBi (CsZnBi) is slightly expanded by 0.23$\%$ (0.78)$\%$ along x-y plane and compressed by 0.92$\%$ (0.23) $\%$ along z-axis, respectively. The reduced interlayer spacing reduces the confinement effect and the band gap of RbZnBi and CsZnBi is decreased to the values of 0.45 eV and 0.44 eV, respectively. 

\begin{figure}[H]
   \centering    
   \includegraphics[width=0.8\textwidth, height = 10cm]{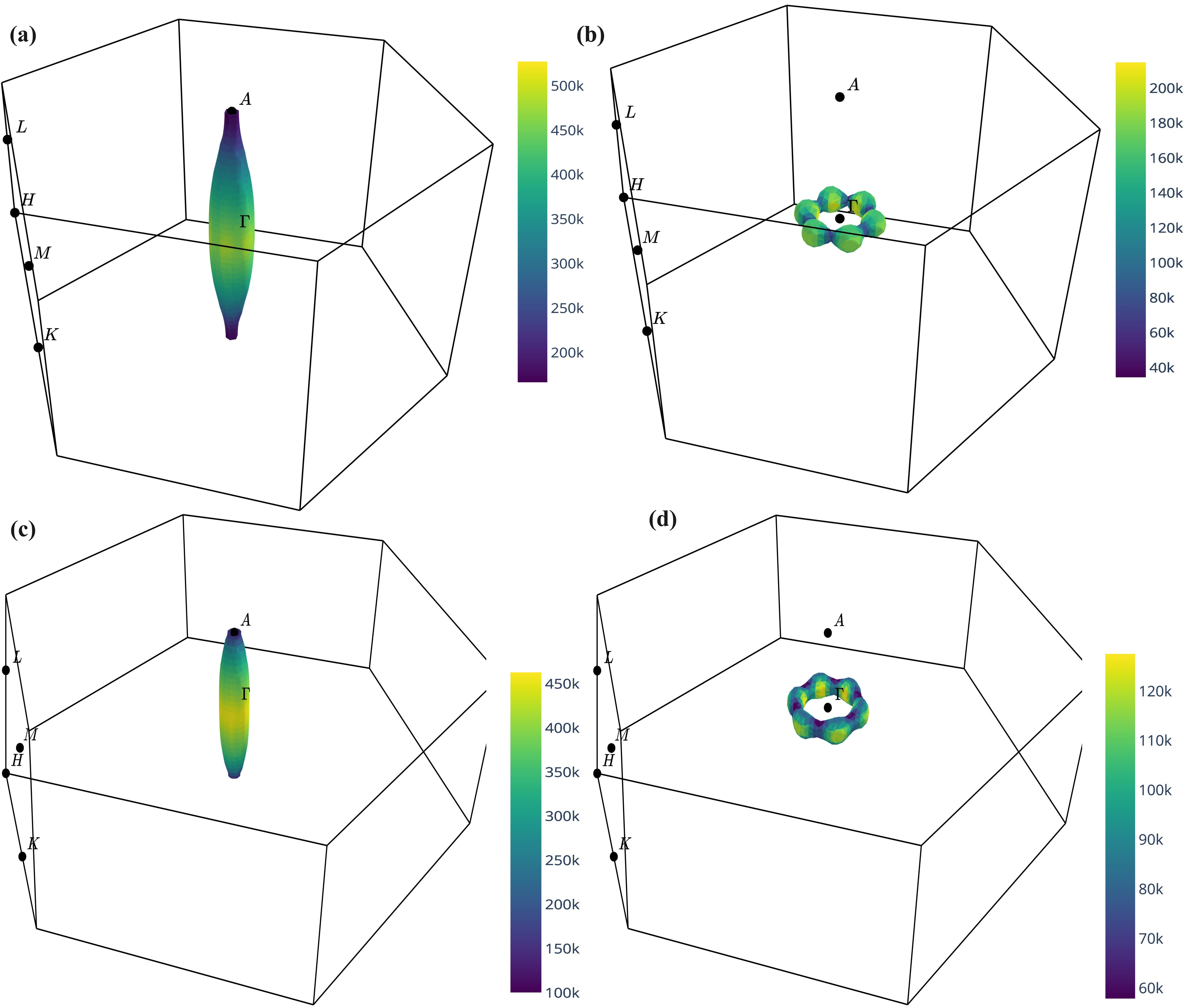}
   \caption{Electron and hole pockets near the conduction band minimum (CBM) and valence band maximum (VBM) for CsZnBi in (a, b) AA-stacking and (c, d) AB stacking. Isoenergy values of 0.35 eV above CBM and 0.22 eV below VBM are used to plot the carrier pockets.}
   \label{fig:1.3}
\end{figure}

The conduction band maximum (CBM) of the studied Zintl-phase materials reveals a distinctive pudding mold-type band structure\cite{adhidewata2022thermoelectric}, characterized by a nearly flat region along the $\Gamma$–A direction and a steeply dispersive segment along $\Gamma$–K. This leads to highly anisotropic effective mass of electrons as shown in Table S1. The unique combination of flat and dispersive features, with anisotropic effective mass, enhances thermoelectric performance by offering both a high density of states (favorable for a large Seebeck coefficient) and high group velocity (beneficial for electrical conductivity)\cite{kuroki2007pudding}. Meanwhile, the valence band edge shows moderate anisotropic dispersion, with a local maximum located along the $\Gamma$–K direction. To further elucidate the dimensionality and symmetry of the electronic structure near the band edges, Fermi isoenergy surfaces are generated using IFermi package\cite{ganose2021ifermi} at energy levels of 0.42 (0.35) eV above the CBM and 0.27 (0.22) below the VBM in AA (AB)-stacking for XZnBi.  Figure \ref{fig:1.3} shows the electron and hole pocketin Figures S2 and Figure for AA and AB stacking, respectively. These energy offsets correspond to thermally active states near room temperature and above, relevant to carrier concentrations of the order of 10$^{19}$ cm$^{-3}$, and effectively capture the dominant transport channels. For AA-stacking, the Fermi surface plots reveal a quasi-two-dimensional, ellipsoidal electron pockets elongated along $k_z$ direction and centered at the $\Gamma$–A line, indicative of anisotropic electron transport. In contrast, the valence band forms a toroidal (ring-shaped) hole pocket centered at the $\Gamma$ point, which is extended within the $k_x$-$k_y$ plane, suggesting enhanced p-type conductivity under hole doping. Due to the underlying hexagonal symmetry of the crystal structure, the valence band maxima (VBM) is replicated along six symmetry-equivalent $\Gamma$–K directions within the Brillouin zone, resulting in the formation of six equivalent hole pockets. These pockets collectively enhance the valence band degeneracy and effectively increase the number of available transport channels for holes. Notably, the conduction band features a single but two-fold degenerate electron pocket at the $\Gamma$ point. This valley degeneracy ($N_v$ = 2) effectively enhances the carrier pocket multiplicity, contributing to a sharp increase in the electronic density of states (DOS) near the band edge. The coexistence of topologically distinct electron and hole pockets, along with pudding mold-like band features, suggests strong potential for bipolar thermoelectric transport in these materials.

For AB-stacked counterparts, the electron pockets are comparatively more compressed and exhibit reduced group velocity contrast, pointing toward enhanced localization of electrons. Similarly, the hole pockets appear flatter, which reflects increased effective mass and reduced contribution to hole conduction. This structural modification due to stacking appears to influence the band curvature and thus the electronic anisotropy. 

\subsection{Electron-Phonon coupling and Relaxation time}
To assess the carrier scattering mechanisms and their impact on transport, the relaxation time ($\tau$) is computed using the EPW code\cite{lee2023electron} at a carrier concentration of 10$^{19}$ cm$^{-3}$, based on first-principles electron–phonon interaction data. The temperature-dependent $\tau$ plots reveal a consistently longer relaxation time for electrons compared to holes over the full temperature range in both the stacking configurations as shown in Figure \ref{fig:1.4} (a). This disparity originates from differences in electronic and phonon band structures and available scattering channels, as described by Fermi’s Golden Rule:
\begin{equation}
\tau^{-1}_{n\mathbf{k}} = \frac{2\pi}{\hbar} \sum_{m,\nu,\mathbf{q}} \left| g^{\nu}_{mn}(\mathbf{k}, \mathbf{q}) \right|^2 \left(1 - f_{m\mathbf{k}+\mathbf{q}} \right) \delta\left( \varepsilon_{n\mathbf{k}} - \varepsilon_{m\mathbf{k}+\mathbf{q}} \pm \hbar \omega_{\nu\mathbf{q}} \right)
\end{equation}
where the scattering rate $\tau^{-1}$ depends on the electron–phonon coupling matrix element $|g^{\nu}{mn}(\mathbf{k}, \mathbf{q})|^2$, the Fermi–Dirac distribution $f$, and the phonon frequency $\omega_{\nu\mathbf{q}}$\cite{gasparov1993electron}. Loosely speaking, the presence of six hole pockets in the valence band provides a dense set of allowed $k \rightarrow k+q$ transitions, enabling frequent intervalley scattering and increasing the number of terms contributing to the scattering summation. Furthermore, the small energy separation between these valleys permits low-energy phonons to easily mediate transitions, enhancing the scattering probability and leading to low $\tau$. Conversely, the conduction band hosts a single electron pocket with limited valley degeneracy and reduced density of states, thereby restricting the number of energetically and momentum-allowed transitions. This suppresses both the matrix element and the phase space for electron–phonon scattering, resulting in longer $\tau$ for electrons than holes. The temperature dependence of $\tau$ primarily stems from the phonon occupation number $n_{\nu\mathbf{q}}$, which increases with temperature and leads to a general trend of $\tau \propto T^{-1.1}$. However, the relaxation time for holes shows a weaker temperature dependence due to their inherently higher initial scattering rates driven by valley multiplicity. Collectively, these findings underscore the interplay between Fermi surface topology, valley degeneracy, and band anisotropy in determining relaxation dynamics and thermoelectric transport behavior.

Comparatively, RbZnBi demonstrates slightly higher relaxation times than CsZnBi for both carrier types, which may stem from differences in their band curvature and band effective masses as shown in Table S1. A flatter band near the Fermi level typically increases the density of states and effective mass, enhancing scattering and thereby reducing $\tau$. Overall, the relaxation time trends observed here reflect the intricate balance between band dispersion, carrier effective mass, scattering channels and phonon coupling strength. These results emphasize that electronic structure engineering, especially tuning the curvature and degeneracy of band edges, plays a pivotal role in optimizing carrier lifetimes and thermoelectric transport.

Importantly, changing the stacking order has a minor effect on the carrier relaxation time. The reason being the electron scattering is predominantly governed by the topology and degeneracy of the Fermi surface, which remain essentially unchanged across different stacking configurations. Hence, the influence of stacking order is minimal on relaxation time and will hardly affect electron transport coefficients.

\subsection{Lattice Dynamics: Thermal conductivity and Associated Parameters}
The macroscopic lattice thermal conductivity ($k_l$) is computed by summing the contributions from all microscopic phonon modes $\lambda(q, j$), where $q$ is the phonon wavevector and $j$ denotes the phonon branch index. The expression used is:
\begin{equation}
          k_{l}  = \frac{1}{N \Omega} \sum_{\lambda} C _{\lambda} (T) v_{\lambda} \otimes v_{\lambda} \tau_{\lambda}(T)
\end{equation}
Here, $\Omega$ is the volume of the unit cell, and $N$ is the total number of sampled $q$-points in the Brillouin zone, which corresponds to the number of unit cells in the crystal\cite{srivastava2022physics}. $C_{\lambda}(T)$ is the mode-dependent heat capacity, $v_{\lambda}$ is the phonon group velocity, and $\tau_{\lambda}(T)$ is the phonon lifetime at temperature $T$. The reciprocal of $\tau_{\lambda}(T)$ is related to the imaginary part of phonon self-energy $(\Gamma _{\lambda} (\omega))$, which takes the form analogous to the Fermi's golden rule\cite{phono3py,phonopy-phono3py-JPCM,PhysRevLett.110.265506,togo2015distributions},
\begin{equation}
\begin{split}
\Gamma_{\lambda}(\omega) = \frac{18\pi}{\hbar^2} \sum_{\lambda' \lambda''} &|\Phi_{-\lambda \lambda' \lambda''}|^2 \{(n_{\lambda'} + n_{\lambda''} + 1) \delta(\omega - \omega_{\lambda'} - \omega_{\lambda''}) \\
&+ (n_{\lambda'} - n_{\lambda''}) [\delta(\omega + \omega_{\lambda'} - \omega_{\lambda''}) - \delta(\omega - \omega_{\lambda'} + \omega_{\lambda''})]\} ,
\end{split}
\end{equation}
The first term in the curly brackets represents collision/decay processes (where one phonon splits into two), while the second term accounts for coalescence processes (where two phonons merge). Also, $\omega$  and $n$ are the phonon frequency and occupation numbers at the equilibrium. The strength of these interactions is governed by third-order interatomic force constants $|\Phi_{-\lambda \lambda' \lambda''}|^2$, where $\lambda, \lambda', \lambda''$ are the wave vectors of three phonons involved in the scattering. This equation shows that the scattering rate is determined by $\Phi_{-\lambda \lambda' \lambda''}$ and the weighted sum over available scattering channels, restricted by energy conservation via the Dirac delta functions $\delta(\omega \pm \omega' \pm \omega'')$.

\begin{figure}[H]
   \centering    
    \includegraphics[width=1\textwidth, height = 12cm]{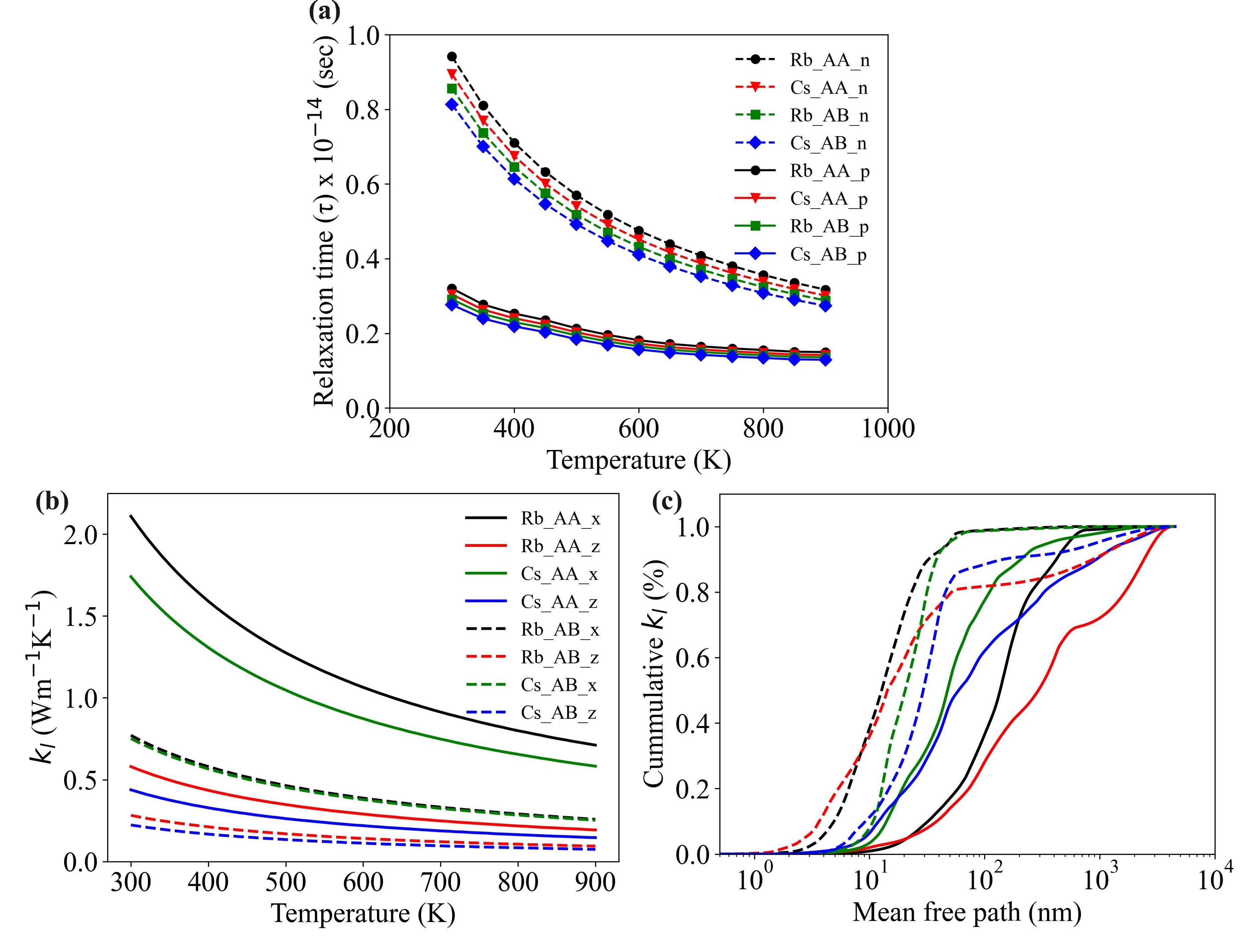}
    \caption{(a) Temperature-dependent electron and hole relaxation time ($\tau$) for XZnBi in AA and AB stacking configurations. The data are obtained at varying Fermi energies but at a fixed carrier concentration (10$^{19}$ cm$^{-3}$). (b) Anisotropic lattice thermal conductivity ($k_{l}$) as a function of temperature along x (z) direction in AA and AB stacked XZnBi. (c) Cumulative lattice thermal conductivity as a function of mean free path along different directions and stacking orders.}
    \label{fig:1.4}
\end{figure}

Figure \ref{fig:1.4} (b) presents the anisotropic $k_{l}$ as a function of temperature along a and c-axis. Both compounds show strong anisotropy in $k_l$, which also varies with the stacking configuration. At room temperature, the $k_{l}$ of AA-stacked RbZnBi and CsZnBi is 2.25 (0.65) Wm$^{-1}$K$^{-1}$ and 1.75 (0.51) Wm$^{-1}$K$^{-1}$ along a (c)-axis, respectively. While in AB-stacking, the room temperature values of $k_{l}$ are 0.96 (0.32) and 0.93 (0.26) Wm$^{-1}$K$^{-1}$ along a (c)-axis, respectively. At 700K, these values further reduce to 1.12 (0.25) Wm$^{-1}$K$^{-1}$ and 0.99 (0.21) Wm$^{-1}$K$^{-1}$ corresponding to Rb and Cs along a (c)-axis, respectively. Interestingly, $k_{l}$ is lower in AB stacked configuration than AA stacked as shown in Table S1. On average, $k_{l}$ \textless  2 Wm$^{-1}$K$^{-1}$ in both the compounds and they can be considered as thermoelectric materials possessing ultra-low lattice thermal conductivity.  The $k_{l}$ is comparable to materials like NbCoSb (~2.5 Wm$^{-1}$K$^{-1}$ for samples synthesized at 900°C)\cite{kumar2024role}, Cs$_{2}$PtI$_{6}$ (0.15 Wm$^{-1}$K$^{-1}$)\cite{sajjad2020ultralow}, Tl$_{3}$VSe$_{4}$ (0.30 Wm$^{-1}$K$^{-1}$ at 300 K)\cite{xia2020particlelike}, Tl$_{2}$O (0.17 Wm$^{-1}$K$^{-1}$ at 300 K)\cite{sajjad2019ultralow}, SnSe along the a-axis (0.7 Wm$^{-1}$K$^{-1}$ at 300 K)\cite{zhao2014ultralow} and in-plane $k_{l}$ of low-temperature thermoelectric material Bi$_{2}$Te$_{3}$ (1.6 Wm$^{-1}$K$^{-1}$ at 300 K)\cite{park2016thermal}. Moreover, the $k_{l}$ values at room temperature are significantly lower than those of conventional ABX thermoelectric materials such as HfCoSb, ZrCoSb, HfZrCoSb, HfZrNiSn (12-18 Wm$^{-1}$K$^{-1}$ at 400 K)\cite{he2018atomistic, he2020unveiling}. The difference in $k_{l}$ in RbZnBi and CsZnBi compounds and its dependancy on stacking order has to do with the difference in dynamical properties associated with $k_{l}$ and is elaborated as follows. 

\begin{figure}[H]
   \centering    
   \includegraphics[width=0.9\textwidth, height = 11cm]{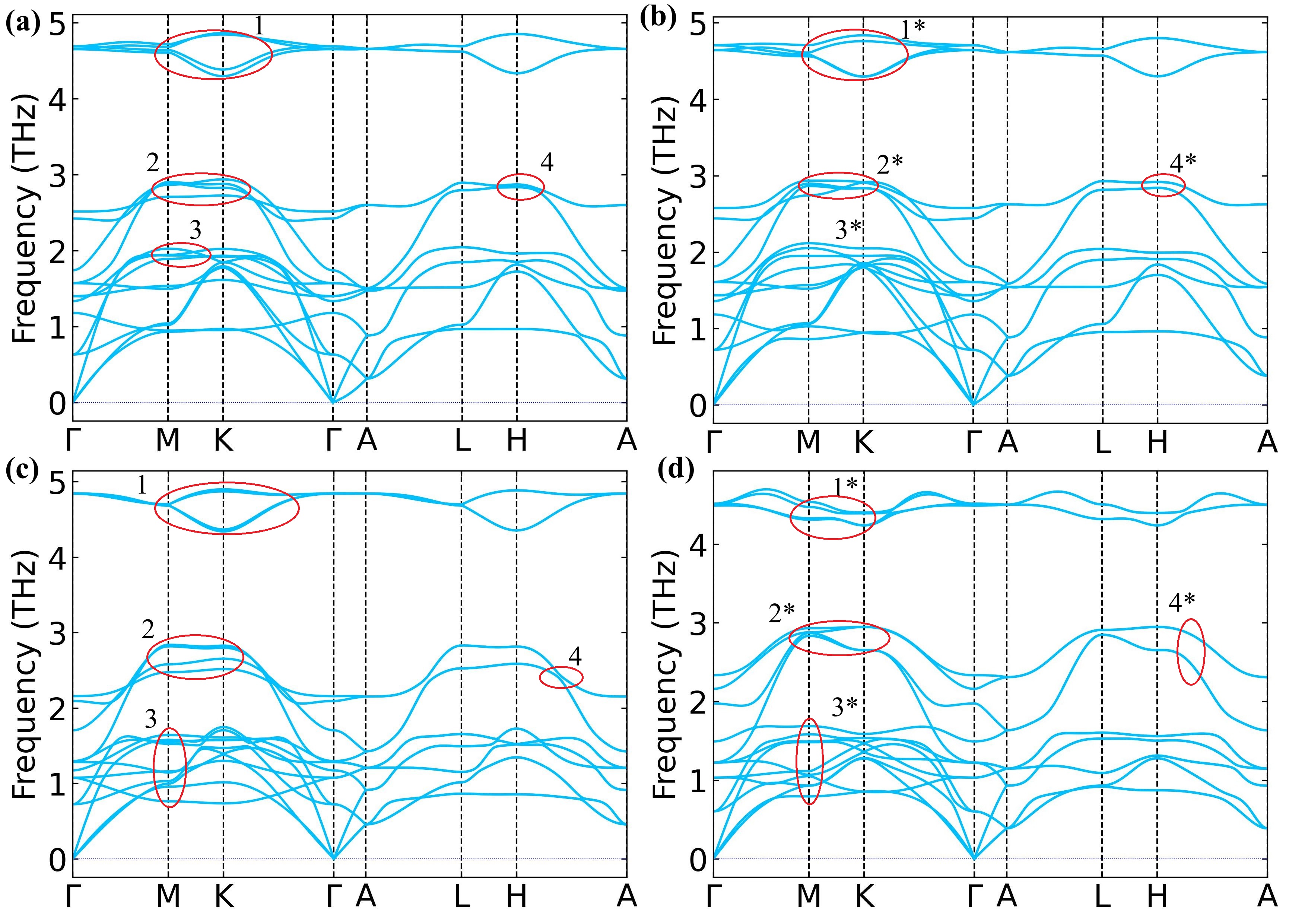}
   \caption{Phonon band structure of (a, b) RbZnBi and (c, d) CsZnBi in AA and AB stacking, respectively. Red circles indicate broken degeneracy due to change in the stacking order. The plots are volume normalized to get a better comparision.}
   \label{fig:1.6}
\end{figure}

The $k_{l}$ of a material can be traced from its phonon dispersion because its behavior is directly associated with the propagation of thermal phonons within a material. Figure \ref{fig:1.6} presents the phonon band structure of RbZnBi (upper row) and CsZnBi (lower row) in AA (a, c) and (AB) (b, d) stacking configurations with a similar number of atoms in the basis. Notably, both materials exhibit a narrow phonon frequency spectrum with optical modes maximizing at a frequency of around 4.6-4.8 THz, which is comparable to some of the high performance thermoelectric materials such as Bi$_{2}$Te$_{3}$, Sb$_{2}$Te$_{3}$, Bi$_{2}$Se$_{3}$ and SnSe\cite{fang2019complex, yang2015enhanced, byun2024simultaneously}. These low phonon frequencies are favorable for minimizing $k_{l}$. On comparing the phonon dispersion at similar high symmetry points, it is found that the degeneracy is lifted at some points in AB stacking order. For instance, the degeneracy at the points in AA stackings, highlighted by red circles and numbered from 1-4, is lifted in AB stacking in both materials marked from 1*-4*. This is a result of the transition from a high-symmetry phase to a low-symmetry one and will lead to an increase in three-phonon selection rules and scattering rates.

To deeply analyze the dynamical behavior of XZnBi compounds, we plotted the projected phonon density of states (PDOS) as shown in Figure S3. The acoustic and low-energy optical phonon modes of XZnSb originate from the contribution of alkali metal (X) and Zn atoms, whereas the high-energy optical phonon modes are primarily associated with Zn and Sb atoms in both stackings. This variation in the contribution of different atoms in different phonon modes is associated with their varying atomic weights. To validate the results of PDOS, we projected the phonon eigen vectors from the dynamical matrix to the real space and computed the corresponding eigen displacements. Figure S4 illustrates the atomic displacements associated with various vibrational modes near the $\Gamma$ point. It is evident that the mid and high-frequency phonon modes are largely governed by the vibrations of Zn and Bi atoms and hence remains largely unaffected by the substitution of X site.  In contrast, X atoms participate predominantly in the acoustic and low-frequency optical modes, exhibiting non-zero displacement amplitudes only in these regimes. These are the modes which dominate heat transportation. Consequently, the substitution of Rb with Cs selectively modifies the phonon spectrum by softening the acoustic and low-lying optical branches. This is reflected in the frequency of longitudinal acoustic (LA) modes, which shifts from approximately 1.82–1.51 THz (in the Rb-containing system) to 1.71–1.25 THz upon Cs incorporation. The observed softening of acoustic modes directly reduces the phonon group velocity as shown in Figure S5.  This reduction leads to a significant suppression of $\kappa_{l}$ in CsZnBi compared to its RbZnBi counterpart. As discussed above, the $\Gamma$–K path, lying in the $x-y$-plane, reflects stronger interatomic bonding and thus exhibits steeper acoustic branches, whereas the $\Gamma$–A direction (along $z$-axis) corresponds to weaker bonding and a softer elastic response. Since phonon group velocity is proportional to the slope of the dispersion, this results in lower group velocity and consequently reduced lattice thermal conductivity along the $z$-direction.  

\begin{figure}[H]
   \centering    
   \includegraphics[width=0.9\textwidth, height = 11cm]{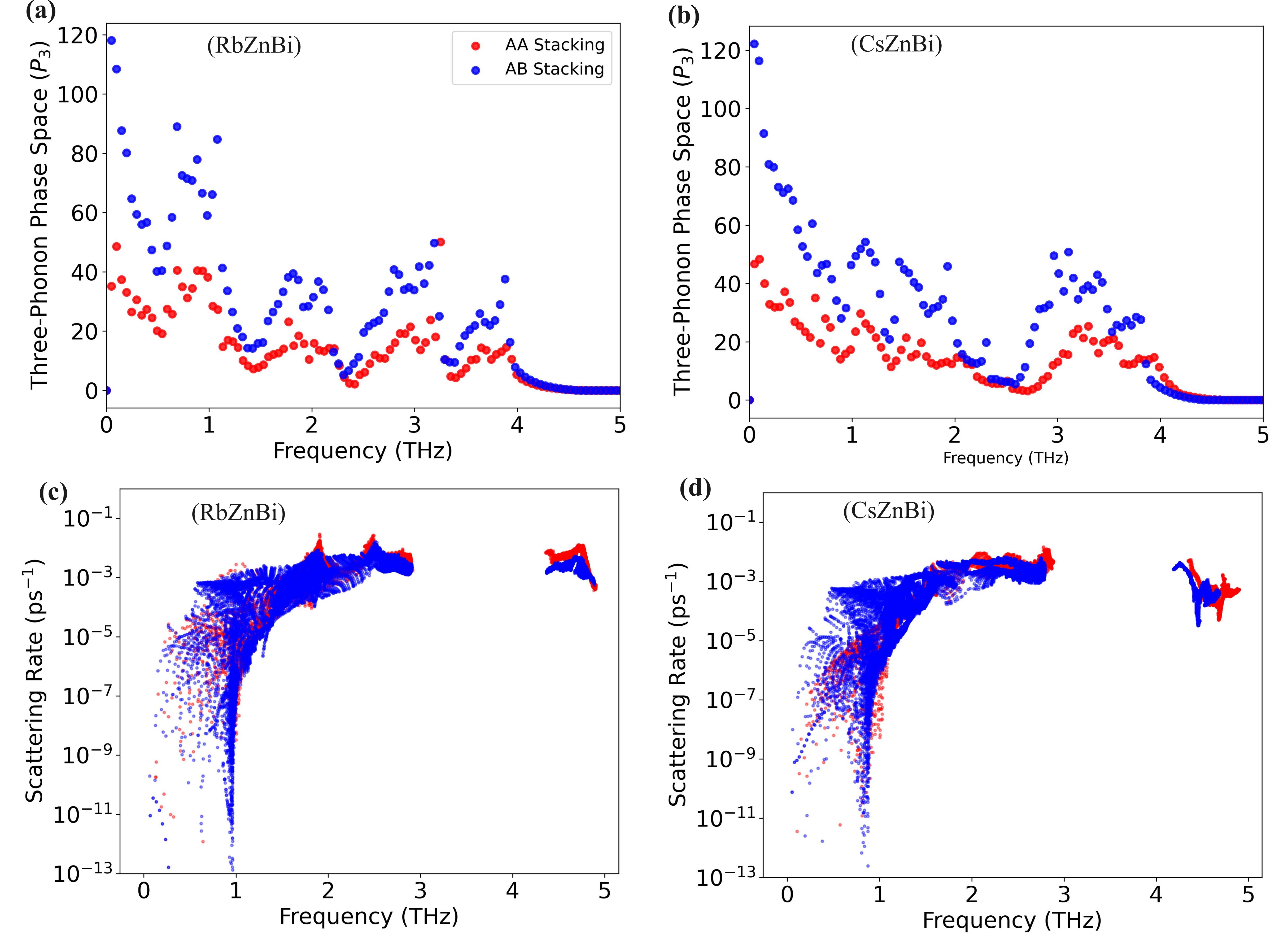}
   \caption{(a, b) Three-phonon phase space (P$_{3}$) and (c, d) phonon scattering rate of RbZnBi and CsZnBi, respectively, in AA and AB-stacking.}
   \label{fig:1.7}
\end{figure}

A substantial disparity in $\kappa_L$  is observed in AA and AB stacking in both the compounds, which is attributed to their distinct vibrational behaviour. Based on the comparative phonon dispersion relations for RbZnBi and CsZnBi, the transition from AA to AB stacking introduces appreciable modifications to the vibrational landscape that directly correlate with increased thermal resistance. To ensure a fair comparison between the AA and AB stacking configurations, all vibrational parameters are normalized relative to the number of atoms and unit cell volume. Dynamically, the AB stacking configuration exhibits a profound enhancement in the three-phonon scattering phase space ($P_3$) as shown in Figure \ref{fig:1.7} (a, b).  This serves as a leading cause of its increased thermal resistance. The structural change in the AB stacking relaxes the crystal selection rules and scattering constraints which were forbidden in AA stacking. This opens up many ways for energy and momentum to be conserved and enhanced inelastic scattering through the $U$-process. The AB configuration displays a nearly three-fold increase in available scattering states within the acoustic frequency regime (below $1.5$ THz). This expansion of the scattering phase space leads to a significant increase in the imaginary part of the phonon self-energy. Physically,  a dramatic increase in phonon scattering rates is observed as shown in Figure \ref{fig:1.7} (c, d). Consequently, the heat-carrying acoustic phonons in the AB lattice undergo significantly more frequent collisions, effectively restricting the thermal transport and providing a clear physical mechanism for the observed suppression of $\kappa_l$.

To provide further physical insight into the suppressed $\kappa_L$ of the AB-stacked phases, we analyzed the mode-resolved Grüneisen parameters ($\gamma$), which serve as a direct gauge of lattice anharmonicity. As illustrated in Figure S6, the AB stacking configuration (blue dots) exhibits a significantly broader distribution and higher absolute magnitudes of $\gamma$ compared to the AA stacking (red dots), particularly in the low-frequency acoustic region ($< 1.5$ THz). This suggests that the structural transition from AA to AB induces a more complex and anharmonic bonding environment. Large positive or negative Grüneisen values indicate that the vibrational modes are highly sensitive to atomic displacements, which physically translates to stronger phonon-phonon interactions. As $\kappa_l$ is inversely proportional to the square of the Grüneisen parameter ($\kappa_l \propto 1/\gamma^2$), the enhanced anharmonicity in the AB phases directly leads to more frequent phonon scattering. While the phase space ($P_3$) defines the number of available scattering channels, the elevated Grüneisen parameters in RbZnBi and CsZnBi indicate that the strength of each scattering event is also intensified in the AB configuration. Consequently, the acoustic phonons experience a much shorter mean free path, thereby solidifying the physical basis for the highly resistive thermal transport and superior thermoelectric potential observed in these AB-stacked materials.

The cumulative lattice thermal conductivity as a function of phonon mean free path (MFP) offers critical insight into phonon transport characteristics and the potential for nanostructuring to suppress $\kappa_{l}$. In both AA and AB stacking configurations of RbZnBi and CsZnBi, a substantial portion of the $\kappa_{l}$ is accumulated from phonons with MFPs below 200 nm as shown in Figure \ref{fig:1.4} (c). This indicates that nanostructures with characteristic dimensions on this scale could effectively scatter dominant heat-carrying phonons. In the AA stacking configuration, the cumulative lattice thermal conductivity of CsZnBi reaches 90$\%$ at a phonon MFP of approximately 100 nm, whereas RbZnBi requires nearly 200 nm to achieve the same thermal conductivity accumulation. This indicates that phonon transport in CsZnBi is more strongly dominated by short-MFP phonons compared to RbZnBi. The intrinsic $\kappa_{l}$ of these materials is low, and MFP analysis suggests a higher potential for $\kappa_{l}$ suppression via nanostructuring at smaller length scales. Similar phonon MFPs are observed in well-known thermoelectric materials such as Bi$_{2}$Te$_{3}$ (1-20 nm)\cite{hellman2014phonon}, PbTe (1-50 nm)\cite{qiu2012molecular} and SnSe (1-100 nm)\cite{xiao2016origin}.  Such observations are highly relevant in the context of experimental strategies for thermal management. For instance, nanoparticle sizes below 10 nm have been successfully realized through solid solution alloying in PbTe-PbS systems, where thermal treatment induces the spontaneous nucleation and growth of PbS nanocrystals\cite{girard2013analysis}. This approach has achieved a remarkable 60$\%$ reduction in the $\kappa_{l}$ of PbTe at 400–500 K\cite{ginting2024optimizing}. Analogously, introducing nanostructures below 20 nm in CsZnBi could offer a promising route to significantly reduce its $\kappa_{l}$, thereby enhancing its thermoelectric performance.

It is important to note that, unlike electronic transport coefficients, which are nearly independent of stacking order, the $\kappa_l$ exhibit a strong dependency on order of stacking. For instance, the $\kappa_l$ in AB-stacking is two times lower than AA-stacked counterpart at room temperature. This is an important feature for achieving decoupling between electronic and thermal transport properties. The reduced $\kappa_l$ in AB stacking originates from selectively enhanced phonon anharmonicity, while the electron-phonon relaxation time remains largely unaffected. Although various strategies have been proposed to decouple electrical and thermal transport, suppression of thermal conductivity often concurrently degrades electrical conductivity, making effective electrical-thermal decoupling highly challenging. In contrast, modifying the stacking order in layered XZnBi intrinsically decouples these transport channels. This demonstrates that layer stacking provides an efficient and physically robust route to electrical-thermal decoupling in these materials.

\subsection{Transport Properties}
\begin{figure}[H]
    \centering    
    \includegraphics[width=1\textwidth, height = 13cm]{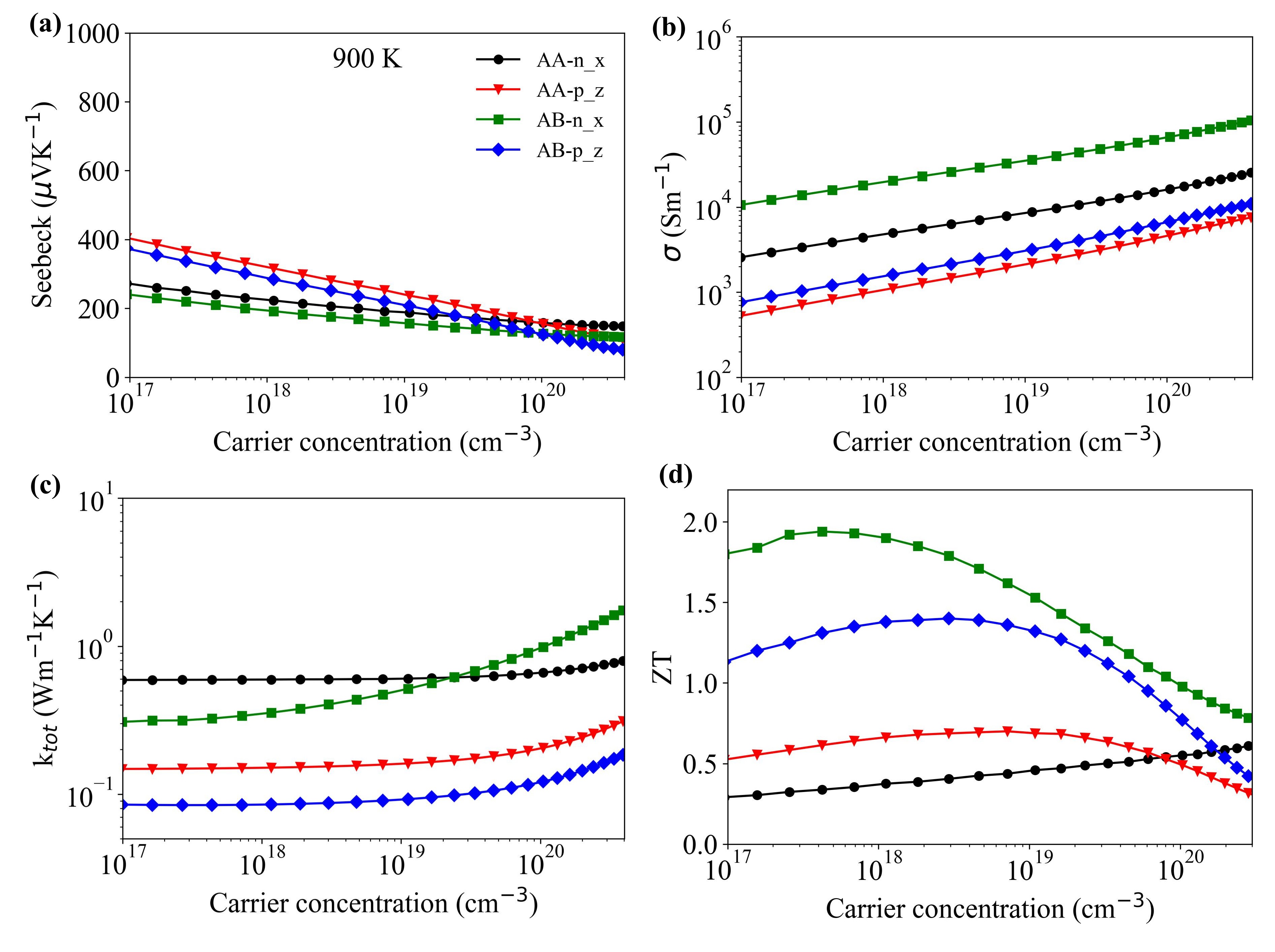}
    \caption{Carrier concentration dependent (a, b) Seebeck coefficient, (c, d) electrical conductivity ($\sigma$), (e, f) total thermal conductivity ($k_{tot}$), and (g, h) Figure of merit (ZT) of CsZnBi at 900 K. Left and right columns represent AA and AB-stacking, respectively. Results are shown for both p-type and n-type doping along x and z-directions.}
    \label{fig:1.8}
 \end{figure}

The transport properties of semiconductors are very sensitive to the doping level or carrier concentration ($n$). A small change in $n$ leads to a large change in transport coefficients. Hence, optimization of $n$ is the easiest approach to tune the thermoelectric parameters of a material.  Figure \ref{fig:1.8} (a) shows the variation of the Seebeck coefficient ($S$) in AA and AB stacking with respect to electron and hole concentration at 900 K. The range of $n$ considered is well within the regime of semiconducting materials. All the phases shows directional dependent $S$ because of the anisotropic electronic structure. 

From Figure \ref{fig:1.8} (a), the magnitude of $S$ exhibits a typical decreasing trend with increasing $n$ across all investigated temperatures (300 K, 600 K, 900 K) in AB stacking of CsZnBi, as expected from Boltzmann transport theory. The same behaviour can be observed at other temperatures and stackings, as shown in Figure S7-S11 (a). At low $n$, the electronic chemical potential is well within the bandgap. In this scenario, the charge carriers observe maximum asymmetry in the density of states and $S$ shows the highest value in this charge concentration. Increasing $n$ pushed the chemical potential inside the filled bands. As a result, carriers observe low asymmetry in density of states and cancel the overall effect of voltage developed by charge transportation. Hence, at higher $n$, $S$ falls monotonically. At low carrier concentrations (10$^{17}$-10$^{18}$ cm$^{-3}$), $S$ reaches values as high as 600 $\mu$VK$^{-1}$ (at 300 K) due to the moderate bandgap of both materials. Notably, RbZnBi consistently shows higher $S$ than CsZnBi for equivalent carrier densities in both stackings, particularly in the p-type x-direction, suggesting better carrier filtering or enhanced effective mass in RbZnBi. Directional anisotropy is also evident for both compounds. The x-direction consistently yields higher $S$ values in the p-type regime, suggesting band anisotropy may play a role in optimizing directional performance. Overall, p-type doping in the z-direction emerges as the most favorable transport channel for maximizing the $S$ in both materials.

Figure \ref{fig:1.8} (b) presents the variation of electrical conductivity ($\sigma$) as a function of $n$ for CsZnBi in AB-stacked configuration, calculated in the range of 300 K-900 K. The plots at other temperatures and stackings are shown in Figure S7-S11 (b). The results are obtained using electron-phonon coupling-based relaxation time derived from the EPW code. In both materials and stacking types, $\sigma$ increases monotonically with increasing $n$, consistent with semiconductor transport behavior. A clear distinction in $\sigma$ arises between single flat bands and multiple dispersive bands. While multiple equivalent valleys enhance the density of states and can increase $\sigma$, they also introduce a higher probability of inter-valley scattering, creating a trade-off between conductivity gain and scattering loss. In contrast, single pudding-mold type bands minimize inter-valley scattering while still providing a high density of states due to their flat regions. Since the calculated $\sigma$ is higher for electrons than for holes in these compounds, the pudding-mold-like conduction bands are identified as more favorable for electrical transport.

The EPW calculations reveal a much more anisotropic behavior of $\sigma$ in both materials. Among the carrier types,  n-type conduction along the x-direction exhibits the highest $\sigma$ values. In AA stacked structures, CsZnBi shows marginally higher $\sigma$ compared to RbZnBi, likely due to its slightly lower band gap. While in AB stacking, the $\sigma$ of both materials becomes comparable due to nearly similar band gaps, suggesting a stacking-induced alteration of the band structure and transport pathways. Importantly, ($\sigma$) exhibits stronger anisotropy across crystallographic directions, with x-direction conductivity generally outperforming z-direction, consistent with layered structure influence. Moreover, EPW calculations account for phonon-limited scattering and temperature-dependent $\tau$, leading to more realistic and experimentally relevant predictions.

Figure \ref{fig:1.8} (c) presents the total thermal conductivity ($k_{tot}$ = $k_{l}$ + $k_{e}$) of AB stacked CsZnBi as a function of $n$ at 900 K. The plots at other temperatures and stacking orders are shown in Figure S7-S11 (c), including RbZnBi. A consistent trend is observed where $k_{tot}$ remains nearly constant at low $n$ and then increases at higher $n$ across all temperatures. The $k_l$ is independent of $n$, while $k_e$ highly depends on $n$ due to its direct proportionality to $\sigma$ according to the Wiedemann–Franz law. This correlation of $k_e$ is especially pronounced at high doping levels, where an increase in carrier density enhances charge carrier mobility and, consequently, heat transport via electrons. Thus, $k_{tot}$ shows an increasing trend at higher $n$. Like $\sigma$, the $k_{tot}$ also exhibits strong directional dependence. The in-plane directions (x) for both n-type and p-type doping possess higher $k_{tot}$ compared to the out-of-plane (z) direction, consistent with the anisotropic nature of the electronic transport seen in $\sigma$ plots. This anisotropy is more significant in the AA stacking than AB stacking, aligning with the more delocalized in-plane orbital contributions in AA-stacked systems.

The close correspondence between the trends in $\sigma$ and $k_{e}$ confirms that electronic thermal transport is tightly coupled with charge transport in these Zintl phases. Therefore, optimization of $\sigma$ via doping or band engineering must also consider its simultaneous effect on $k_{e}$, to achieve a high thermoelectric figure of merit ($ZT$). These results emphasize the necessity of energy-dependent relaxation time models such as EPW to avoid overestimation of $k_{e}$ and $\sigma$ in predictive thermoelectric design.

\begin{figure}[H]
    \centering    
    \includegraphics[width=\textwidth, height = 7cm]{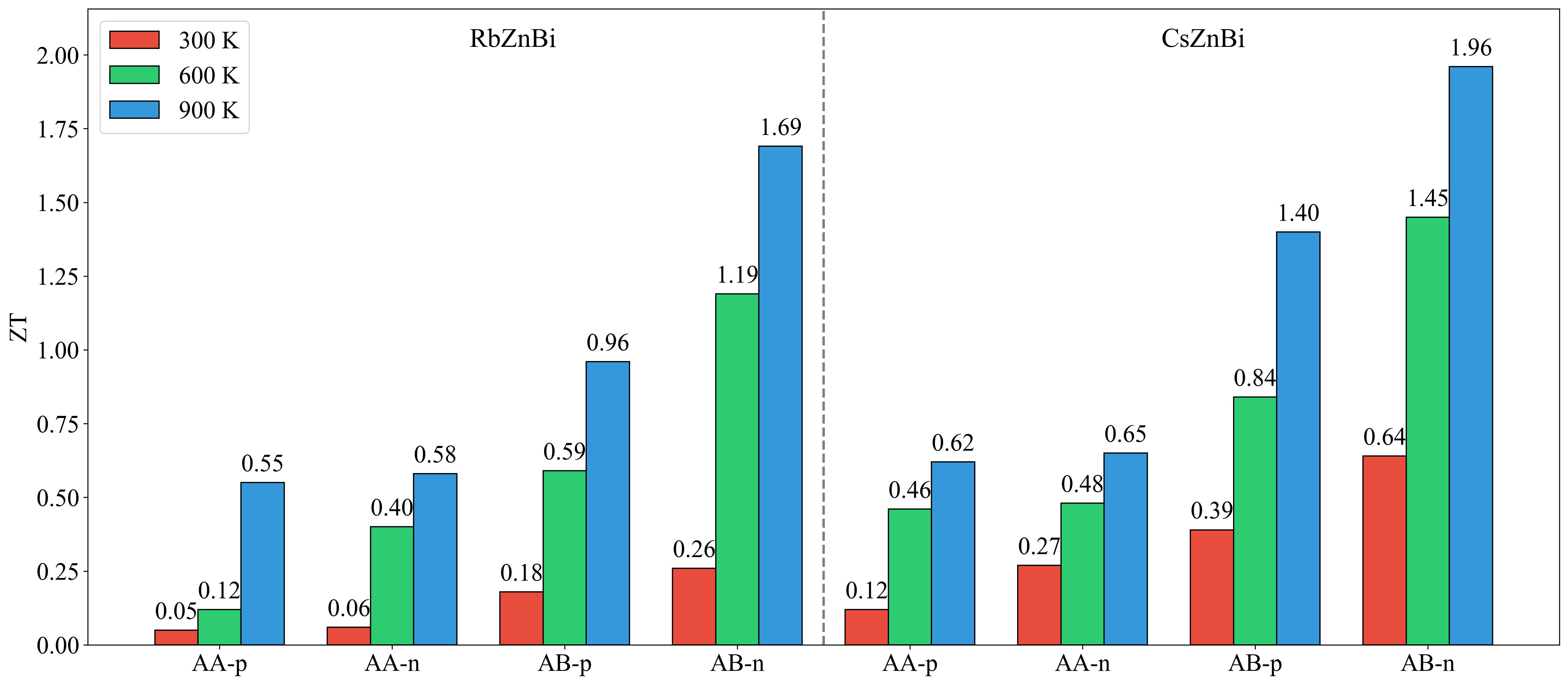}
    \caption{Figure of merit ($ZT$) comparison of n and p-type (x-direction) XZnBi in AA and AB-stacking at different temperatures. The values mentioned are at optimum carrier concentrations.}
    \label{fig:2.2}
\end{figure}

The figure of merit ($ZT$), which encapsulates the overall thermoelectric efficiency of a material, is presented in Figure \ref{fig:1.8} (d) for AB-stacked CsZnBi. Figure S7-S11 (d) demonstrated the ZT at other temperatures and staking orders. To evaluate the thermoelectric performance, the $ZT$ is calculated across a wide range of $n$ and three operational temperatures (300 K-900 K). For each specific $n$, the corresponding Fermi energy was determined to evaluate the electronic transport coefficients, allowing us to identify the optimal Fermi level that maximizes $ZT$ at each temperature. The $ZT$ reveals that AB-stacked configurations consistently outperform their AA-stacked counterparts across all studied $n$ and temperatures, as shown in Figure \ref{fig:2.2}. The core of this higher performance lies in a unique electrical-thermal decoupling mechanism. While both stacking orders exhibit nearly identical electronic transport profiles, the thermal transport in the AB-stacking arrangement is characterized by high thermal resistivity. This suppression of thermal conductivity leads to $ZT$ values for AB-stacking that are more than 40 $\%$  higher than those of the AA-stacking. The doubling of the unit cell in the AB configuration naturally increases the number of available scattering channels. However, the observed increase in $ZT$ is not a numerical artifact of cell size, but the low symmetry in the AB structure, which significantly enhances the anharmonicity and creates a more congested scattering phase space ($P_3$) that is intrinsically more resistive to heat flow than the AA configuration. Notably, the disparity in relaxation times between charge carriers directly influences the doping-dependent $ZT$ peaks. Since electrons exhibit significantly longer relaxation times than holes, the $ZT$ reaches its maxima under $n$-type (electron) doping, while remaining comparatively lower for $p$-type (hole) doping. The peak value of $ZT$ is 1.99 (1.69) for AB-stacked CsZnBi (RbZnBi) with hole doping. The high values put these materials in the category of theoretically studied state-of-the-art thermoelectric materials. These results suggest that optimizing the electrical-thermal coupling is a potential strategy for designing high-efficiency thermoelectric materials by targeting thermal resistance without sacrificing electronic mobility.

\section{Conclusion}
In summary, we have evaluated the thermoelectric properties of the layered Zintl compounds RbZnBi and CsZnBi, establishing stacking order as a decisive mechanism for achieving electrical-thermal decoupling. While electronic transport remains largely insensitive to stacking order due to the preserved Fermi-surface topology, the lattice thermal conductivity ($\kappa_l$) is dramatically suppressed in AB-stacked phases. Our analysis reveals that this suppression originates from a reduction in crystal symmetry, which lifts vibrational degeneracies and relaxes the umklapp scattering selection rules. This significantly enhances the three-phonon scattering phase space ($P_3$) and scattering rates, which makes AB stacked configurations more thermal resistive than AA stacking. This significantly reduces $\kappa_l$ in AB-stacked phase without a proportional effect on electronic transport, leading to a substantial enhancement in the thermoelectric figure of merit. Collectively, these findings identify XZnBi materials as promising thermoelectric candidates and highlight stacking-controlled phonon transport as a robust strategy for the design of advanced thermoelectric materials.

\section*{CRediT authorship contribution statement}
\textbf{Aadil Fayaz Wani:} Conceptualization; Data curation; Investigation; Visualization; Writing - original draft; Writing - review and editing. \textbf{Nirma Kumari:} Conceptualization; Visualization; Writing - review and editing. \textbf{SuDong Park:} Conceptualization; Funding acquisition; Project administration; Writing - review and editing, \textbf{Byungki Ryu:} Conceptualization; Formal analysis; Validation; Funding acquisition; Project administration; Writing - review and editing.

\section*{Acknowledgments}
This work was supported by the National Research Foundation of Korea (NRF) grant funded by the Korean Government (MSIT) (Grant No. 2022M3C1C8093916, which has been assigned a new number, RS-2022-NR119739). It is also was supported by the Primary Research Program of KERI through the National Research Council of Science and Technology (NST) funded by the Ministry of Science and ICT (MSIT) (Grant No. 26A01005) and by the Korea Institute of Energy Technology Evaluation and Planning (KETEP) grant funded by the Ministry of Trade, Industry and Energy (MOTIE) (Grant No. 2021202080023D, which has been assigned a new number, RS-2021-KP002416). It was also partially supported by a National Research Foundation of Korea (NRF) grant funded by the Korean Government (MSIT) (Grant No. RS-2025-25461730).

\section*{Declaration of competing interest}
The authors declare that they have no known competing financial interests or personal relationships that could have appeared to influence the work reported in this paper.

\section*{Data availability}
Data will be made available on request.

\section*{Figures}

\printbibliography

@article{bell2008cooling,
  title={Cooling, heating, generating power, and recovering waste heat with thermoelectric systems},
  author={Bell, Lon E},
  journal={science},
  volume={321},
  number={5895},
  pages={1457--1461},
  year={2008},
  publisher={American Association for the Advancement of Science}
}

@article{hsu2004cubic,
  title={Cubic AgPb m SbTe2+ m: bulk thermoelectric materials with high figure of merit},
  author={Hsu, Kuei Fang and Loo, Sim and Guo, Fu and Chen, Wei and Dyck, Jeffrey S and Uher, Ctirad and Hogan, Tim and Polychroniadis, E Kanatzidis and Kanatzidis, Mercouri G},
  journal={Science},
  volume={303},
  number={5659},
  pages={818--821},
  year={2004},
  publisher={American Association for the Advancement of Science}
}

@article{snyder2008complex,
  title={Complex thermoelectric materials},
  author={Snyder, G Jeffrey and Toberer, Eric S},
  journal={Nature materials},
  volume={7},
  number={2},
  pages={105--114},
  year={2008},
  publisher={Nature Publishing Group UK London}
}

@article{xiao2014decoupling,
  title={Decoupling interrelated parameters for designing high performance thermoelectric materials},
  author={Xiao, Chong and Li, Zhou and Li, Kun and Huang, Pengcheng and Xie, Yi},
  journal={Accounts of chemical research},
  volume={47},
  number={4},
  pages={1287--1295},
  year={2014},
  publisher={ACS Publications}
}

@article{singh2024advancements,
  title={Advancements in thermoelectric materials for efficient waste heat recovery and renewable energy generation},
  author={Singh, Rakesh and Dogra, Surya and Dixit, Saurav and Vatin, Nikolai Ivanovich and Bhardwaj, Rajesh and Sundramoorthy, Ashok K and Perera, HCS and Patole, Shashikant P and Mishra, Rajneesh Kumar and Arya, Sandeep},
  journal={Hybrid Advances},
  volume={5},
  pages={100176},
  year={2024},
  publisher={Elsevier}
}

@inproceedings{weidenkaff2017thermoelectricity,
  title={Thermoelectricity for future sustainable energy technologies},
  author={Weidenkaff, Anke},
  booktitle={EPJ Web of Conferences},
  volume={148},
  pages={00010},
  year={2017},
  organization={EDP Sciences}
}

@article{song2025reduced,
  title={Reduced thermal conductivity for enhanced thermoelectric energy conversion efficiency in p-type (Bi, Sb) 2Te3},
  author={Song, Mingzhen and Zhang, Fudong and Jia, Beiquan and Wang, Weishuai and Shi, Yalin and Peng, Zhanhui and Chao, Xiaolian and Yang, Zupei and Wu, Di},
  journal={Ceramics International},
  year={2025},
  publisher={Elsevier}
}

@article{zhao2011toward,
  title={Toward high-performance nanostructured thermoelectric materials: the progress of bottom-up solution chemistry approaches},
  author={Zhao, Yixin and Dyck, Jeffrey S and Burda, Clemens},
  journal={Journal of Materials Chemistry},
  volume={21},
  number={43},
  pages={17049--17058},
  year={2011},
  publisher={Royal Society of Chemistry}
}

@article{wu2013origin,
  title={Origin of phonon glass--electron crystal behavior in thermoelectric layered cobaltate},
  author={Wu, Lijun and Meng, Qingping and Jooss, Christian and Zheng, Jin-Cheng and Inada, H and Su, Dong and Li, Qiang and Zhu, Yimei},
  journal={Advanced Functional Materials},
  volume={23},
  number={46},
  pages={5728--5736},
  year={2013},
  publisher={Wiley Online Library}
}

@article{zhu2018thermoelectric,
  title={Thermoelectric phonon-glass electron-crystal via ion beam patterning of silicon},
  author={Zhu, Taishan and Swaminathan-Gopalan, Krishnan and Stephani, Kelly and Ertekin, Elif},
  journal={Physical Review B},
  volume={97},
  number={17},
  pages={174201},
  year={2018},
  publisher={APS}
}

@article{wei2023rattling,
  title={Rattling vibrations and occupied antibonding states yield intrinsically low thermal conductivity of the Zintl-phase compound KSrBi},
  author={Wei, Congying and Feng, Zhenzhen and Yan, Yuli and Zhao, Gaofeng and Fu, Yuhao and Singh, David J},
  journal={Physical Review B},
  volume={108},
  number={23},
  pages={235203},
  year={2023},
  publisher={APS}
}

@article{lee2024large,
  title={Large-Gap Z 2 and Topological Crystalline Insulating Phase in RbZnBi and CsZnBi},
  author={Lee, Hyunggeun and Han, Myung Joon and Chang, Kee Joo},
  journal={ACS omega},
  volume={9},
  number={27},
  pages={29820--29828},
  year={2024},
  publisher={ACS Publications}
}

@article{schoop2018chemical,
  title={Chemical principles of topological semimetals},
  author={Schoop, Leslie M and Pielnhofer, Florian and Lotsch, Bettina V},
  journal={Chemistry of Materials},
  volume={30},
  number={10},
  pages={3155--3176},
  year={2018},
  publisher={ACS Publications}
}

@article{owens2020synthesis,
  title={Synthesis, crystal and electronic structure of layered AMSb compounds (A= Rb, Cs; M= Zn, Cd)},
  author={Owens-Baird, Bryan and Wang, Lin-Lin and Lee, Shannon and Kovnir, Kirill},
  journal={Zeitschrift f{\"u}r anorganische und allgemeine Chemie},
  volume={646},
  number={14},
  pages={1079--1085},
  year={2020},
  publisher={Wiley Online Library}
}

@article{xu2015two,
  title={Two-dimensional oxide topological insulator with iron-pnictide superconductor LiFeAs structure},
  author={Xu, Qiunan and Song, Zhida and Nie, Simin and Weng, Hongming and Fang, Zhong and Dai, Xi},
  journal={Physical Review B},
  volume={92},
  number={20},
  pages={205310},
  year={2015},
  publisher={APS}
}

@article{khireddine2021first,
  title={First-principles predictions of the structural, electronic, optical and elastic properties of the zintl-phases AE3GaAs3 (AE= Sr, Ba)},
  author={Khireddine, A and Bouhemadou, A and Alnujaim, S and Guechi, N and Bin-Omran, S and Al-Douri, Y and Khenata, R and Maabed, S and Kushwaha, AK},
  journal={Solid State Sciences},
  volume={114},
  pages={106563},
  year={2021},
  publisher={Elsevier}
}

@article{kauzlarich2023zintl,
  title={Zintl phases: From curiosities to impactful materials},
  author={Kauzlarich, Susan M},
  journal={Chemistry of Materials},
  volume={35},
  number={18},
  pages={7355--7362},
  year={2023},
  publisher={ACS Publications}
}

@article{wu2025zintl,
  title={The Zintl-Klemm Concept in the Amorphous State: A Case Study of Na-P Battery Anodes},
  author={Wu, Litong and Deringer, Volker L},
  journal={arXiv preprint arXiv:2504.04920},
  year={2025}
}

@article{zhang2025anharmonic,
  title={Anharmonic Rattling Vibrations and Multivalley Bands in enabling Zintl-phase YbZn2X2 (X= As, Sb) Thermoelectrics},
  author={Zhang, Zhiwei and Tang, Shuwei and Bai, Shulin and Wan, Da and Ai, Peng and Zhang, Pengfei and Xu, Zhanpeng and Bao, Yujie and Zhang, Yunzhuo},
  journal={Materials Today Physics},
  pages={101837},
  year={2025},
  publisher={Elsevier}
}

@article{liu2019recent,
  title={Recent progresses on thermoelectric Zintl phases: Structures, materials and optimization},
  author={Liu, Ke-Feng and Xia, Sheng-Qing},
  journal={Journal of Solid State Chemistry},
  volume={270},
  pages={252--264},
  year={2019},
  publisher={Elsevier}
}

@incollection{nolas2001phonon,
  title={The phonon—glass electron-crystal approach to thermoelectric materials research},
  author={Nolas, George S and Sharp, Jeffrey and Goldsmid, H Julian},
  booktitle={Thermoelectrics: Basic Principles and New Materials Developments},
  pages={177--207},
  year={2001},
  publisher={Springer}
}

@article{kresse1996efficient,
  title={Efficient iterative schemes for ab initio total-energy calculations using a plane-wave basis set},
  author={Kresse, Georg and Furthm{\"u}ller, J{\"u}rgen},
  journal={Physical review B},
  volume={54},
  number={16},
  pages={11169},
  year={1996},
  publisher={APS}
}

@article{giannozzi2009quantum,
  title={QUANTUM ESPRESSO: a modular and open-source software project for quantumsimulations of materials},
  author={Giannozzi, Paolo and Baroni, Stefano and Bonini, Nicola and Calandra, Matteo and Car, Roberto and Cavazzoni, Carlo and Ceresoli, Davide and Chiarotti, Guido L and Cococcioni, Matteo and Dabo, Ismaila and others},
  journal={Journal of physics: Condensed matter},
  volume={21},
  number={39},
  pages={395502},
  year={2009},
  publisher={IOP Publishing}
}

@article{perdew1996generalized,
  title={Generalized gradient approximation made simple},
  author={Perdew, John P and Burke, Kieron and Ernzerhof, Matthias},
  journal={Physical review letters},
  volume={77},
  number={18},
  pages={3865},
  year={1996},
  publisher={APS}
}

@article{pack1977special,
  title={" Special points for Brillouin-zone integrations"—a reply},
  author={Pack, James D and Monkhorst, Hendrik J},
  journal={Physical Review B},
  volume={16},
  number={4},
  pages={1748},
  year={1977},
  publisher={APS}
}

@article{madsen2018boltztrap2,
  title={BoltzTraP2, a program for interpolating band structures and calculating semi-classical transport coefficients},
  author={Madsen, Georg KH and Carrete, Jes{\'u}s and Verstraete, Matthieu J},
  journal={Computer Physics Communications},
  volume={231},
  pages={140--145},
  year={2018},
  publisher={Elsevier}
}

@article{lee2023electron,
  title={Electron--phonon physics from first principles using the EPW code},
  author={Lee, Hyungjun and Ponc{\'e}, Samuel and Bushick, Kyle and Hajinazar, Samad and Lafuente-Bartolome, Jon and Leveillee, Joshua and Lian, Chao and Lihm, Jae-Mo and Macheda, Francesco and Mori, Hitoshi and others},
  journal={npj Computational Materials},
  volume={9},
  number={1},
  pages={156},
  year={2023},
  publisher={Nature Publishing Group UK London}
}

@article{sjakste2015wannier,
  title={Wannier interpolation of the electron-phonon matrix elements in polar semiconductors: Polar-optical coupling in GaAs},
  author={Sjakste, J and Vast, N and Calandra, M and Mauri, FRANCESCO},
  journal={Physical Review B},
  volume={92},
  number={5},
  pages={054307},
  year={2015},
  publisher={APS}
}

@book{ziman2001electrons,
  title={Electrons and phonons: the theory of transport phenomena in solids},
  author={Ziman, John M},
  year={2001},
  publisher={Oxford university press}
}

@article{togo2015distributions,
  title={Distributions of phonon lifetimes in Brillouin zones},
  author={Togo, Atsushi and Chaput, Laurent and Tanaka, Isao},
  journal={Physical review B},
  volume={91},
  number={9},
  pages={094306},
  year={2015},
  publisher={APS}
}

@article{guyer1966solution,
  title={Solution of the linearized phonon Boltzmann equation},
  author={Guyer, Robert Alan and Krumhansl, JA},
  journal={Physical Review},
  volume={148},
  number={2},
  pages={766},
  year={1966},
  publisher={APS}
}

@article{kim2015characterization,
  title={Characterization of Lorenz number with Seebeck coefficient measurement},
  author={Kim, Hyun-Sik and Gibbs, Zachary M and Tang, Yinglu and Wang, Heng and Snyder, G Jeffrey},
  journal={APL materials},
  volume={3},
  number={4},
  year={2015},
  publisher={AIP Publishing}
}

@article{togo2023implementation,
  title={Implementation strategies in phonopy and phono3py},
  author={Togo, Atsushi and Chaput, Laurent and Tadano, Terumasa and Tanaka, Isao},
  journal={Journal of Physics: Condensed Matter},
  volume={35},
  number={35},
  pages={353001},
  year={2023},
  publisher={IOP Publishing}
}

@article{matar2019coloring,
  title={Coloring in the ZrBeSi-type structure},
  author={Matar, Samir F and P{\"o}ttgen, Rainer},
  journal={Zeitschrift f{\"u}r Naturforschung B},
  volume={74},
  number={4},
  pages={307--318},
  year={2019},
  publisher={De Gruyter}
}

@article{nesper2014zintl,
  title={The Zintl-Klemm concept--a historical survey},
  author={Nesper, Reinhard},
  journal={Zeitschrift f{\"u}r anorganische und allgemeine Chemie},
  volume={640},
  number={14},
  pages={2639--2648},
  year={2014},
  publisher={Wiley Online Library}
}

@article{chesnut1999electron,
  title={The electron localization function (ELF) description of the PO bond in phosphine oxide},
  author={Chesnut, DB and Savin, A},
  journal={Journal of the American Chemical Society},
  volume={121},
  number={10},
  pages={2335--2336},
  year={1999},
  publisher={American Chemical Society}
}

@article{gasparov1993electron,
  title={Electron-phonon, electron-electron and electron-surface scattering in metals from ballistic effects},
  author={Gasparov, VA and Huguenin, R},
  journal={Advances in Physics},
  volume={42},
  number={4},
  pages={393--521},
  year={1993},
  publisher={Taylor \& Francis}
}

@article{kumar2024role,
  title={Role of Nb vacancies and Sn substitution in modulating the thermoelectric properties of NbCoSb},
  author={Kumar, Inder and Peter, Jipin and Shankar, Gyan and Pambannan, Padaikathan and Suwas, Satyam and Biswas, Raju K and Mallik, Ramesh Chandra},
  journal={Physical Review B},
  volume={110},
  number={20},
  pages={205207},
  year={2024},
  publisher={APS}
}

@article{sajjad2020ultralow,
  title={Ultralow lattice thermal conductivity in double perovskite Cs2PtI6: a promising thermoelectric material},
  author={Sajjad, Muhammad and Mahmood, Qasim and Singh, Nirpendra and Larsson, J Andreas},
  journal={ACS applied energy Materials},
  volume={3},
  number={11},
  pages={11293--11299},
  year={2020},
  publisher={ACS Publications}
}

@article{xia2020particlelike,
  title={Particlelike phonon propagation dominates ultralow lattice thermal conductivity in crystalline Tl 3 VSe 4},
  author={Xia, Yi and Pal, Koushik and He, Jiangang and Ozoli{\c{n}}{\v{s}}, Vidvuds and Wolverton, Chris},
  journal={Physical Review Letters},
  volume={124},
  number={6},
  pages={065901},
  year={2020},
  publisher={APS}
}

@article{sajjad2019ultralow,
  title={Ultralow lattice thermal conductivity and thermoelectric properties of monolayer Tl2O},
  author={Sajjad, Muhammad and Singh, Nirpendra and Sattar, Shahid and De Wolf, Stefaan and Schwingenschlogl, Udo},
  journal={ACS Applied Energy Materials},
  volume={2},
  number={5},
  pages={3004--3008},
  year={2019},
  publisher={ACS Publications}
}

@article{zhao2014ultralow,
  title={Ultralow thermal conductivity and high thermoelectric figure of merit in SnSe crystals},
  author={Zhao, Li-Dong and Lo, Shih-Han and Zhang, Yongsheng and Sun, Hui and Tan, Gangjian and Uher, Ctirad and Wolverton, Christopher and Dravid, Vinayak P and Kanatzidis, Mercouri G},
  journal={nature},
  volume={508},
  number={7496},
  pages={373--377},
  year={2014},
  publisher={Nature Publishing Group UK London}
}

@article{park2016thermal,
  title={Thermal and electrical conduction of single-crystal Bi2Te3 nanostructures grown using a one step process},
  author={Park, Dambi and Park, Sungjin and Jeong, Kwangsik and Jeong, Hong-Sik and Song, Jea Yong and Cho, Mann--Ho},
  journal={Scientific reports},
  volume={6},
  number={1},
  pages={19132},
  year={2016},
  publisher={Nature Publishing Group UK London}
}

@article{he2018atomistic,
  title={Atomistic study of the electronic contact resistivity between the half-Heusler alloys (HfCoSb, HfZrCoSb, HfZrNiSn) and the metal Ag},
  author={He, Yuping and L{\'e}onard, Fran{\c{c}}ois and Spataru, Catalin D},
  journal={Physical Review Materials},
  volume={2},
  number={6},
  pages={065401},
  year={2018},
  publisher={APS}
}

@article{he2020unveiling,
  title={Unveiling the phonon scattering mechanisms in half-Heusler thermoelectric compounds},
  author={He, Ran and Zhu, Taishan and Wang, Yumei and Wolff, Ulrike and Jaud, Jean-Christophe and Sotnikov, Andrei and Potapov, Pavel and Wolf, Daniel and Ying, Pingjun and Wood, Max and others},
  journal={Energy \& Environmental Science},
  volume={13},
  number={12},
  pages={5165--5176},
  year={2020},
  publisher={Royal Society of Chemistry}
}

@article{fang2019complex,
  title={Complex band structures and lattice dynamics of Bi2Te3-based compounds and solid solutions},
  author={Fang, Teng and Li, Xin and Hu, Chaoliang and Zhang, Qi and Yang, Jiong and Zhang, Wenqing and Zhao, Xinbing and Singh, David J and Zhu, Tiejun},
  journal={Advanced Functional Materials},
  volume={29},
  number={28},
  pages={1900677},
  year={2019},
  publisher={Wiley Online Library}
}

@article{yang2015enhanced,
  title={Enhanced thermoelectric performance of nanostructured Bi2Te3 through significant phonon scattering},
  author={Yang, Lei and Chen, Zhi-Gang and Hong, Min and Han, Guang and Zou, Jin},
  journal={ACS applied materials \& interfaces},
  volume={7},
  number={42},
  pages={23694--23699},
  year={2015},
  publisher={ACS Publications}
}

@article{byun2024simultaneously,
  title={Simultaneously engineering electronic and phonon band structures for high-performance n-type polycrystalline SnSe},
  author={Byun, Sejin and Ge, Bangzhi and Song, Hyungjun and Cho, Sung-Pyo and Hong, Moo Sun and Im, Jino and Chung, In},
  journal={Joule},
  volume={8},
  number={5},
  pages={1520--1538},
  year={2024},
  publisher={Elsevier}
}

@article{girard2013analysis,
  title={Analysis of phase separation in high performance PbTe--PbS thermoelectric materials},
  author={Girard, Steven N and Schmidt-Rohr, Klaus and Chasapis, Thomas C and Hatzikraniotis, Euripides and Njegic, Bosiljka and Levin, EM and Rawal, Aditya and Paraskevopoulos, Konstantinos M and Kanatzidis, Mercouri G},
  journal={Advanced Functional Materials},
  volume={23},
  number={6},
  pages={747--757},
  year={2013},
  publisher={Wiley Online Library}
}

@article{hellman2014phonon,
  title={Phonon thermal transport in Bi 2 Te 3 from first principles},
  author={Hellman, Olle and Broido, David A},
  journal={Physical Review B},
  volume={90},
  number={13},
  pages={134309},
  year={2014},
  publisher={APS}
}

@inproceedings{qiu2012molecular,
  title={Molecular dynamics simulations of lattice thermal conductivity and spectral phonon mean free path of PbTe: Bulk and nanostructures},
  author={Qiu, Bo and Bao, Hua and Ruan, Xiulin and Zhang, Genqiang and Wu, Yue},
  booktitle={Heat Transfer Summer Conference},
  volume={44779},
  pages={659--670},
  year={2012},
  organization={American Society of Mechanical Engineers}
}

@article{xiao2016origin,
  title={Origin of low thermal conductivity in SnSe},
  author={Xiao, Yu and Chang, Cheng and Pei, Yanling and Wu, Di and Peng, Kunling and Zhou, Xiaoyuan and Gong, Shengkai and He, Jiaqing and Zhang, Yongsheng and Zeng, Zhi and others},
  journal={Physical Review B},
  volume={94},
  number={12},
  pages={125203},
  year={2016},
  publisher={APS}
}

@incollection{ginting2024optimizing,
  title={Optimizing Thermal Conductivity in PbTe: Nanocomposite and Alloy Approaches for Low Thermal Conductivity},
  author={Ginting, Dianta and Rhyee, Jong-Soo},
  booktitle={Current Research in Thermal Conductivity},
  year={2024},
  publisher={IntechOpen}
}

@article{kuroki2007pudding,
  title={“Pudding mold” band drives large thermopower in NaxCoO2},
  author={Kuroki, Kazuhiko and Arita, Ryotaro},
  journal={Journal of the Physical Society of Japan},
  volume={76},
  number={8},
  pages={083707--083707},
  year={2007},
  publisher={The Physical Society of Japan}
}

@article{mk2023layer,
  title={Layer number and stacking order dependent thermal transport in molybdenum disulfide with sulfur vacancies},
  author={MK, Ranjuna and Balakrishnan, Jayakumar},
  journal={Physical Review B},
  volume={108},
  number={24},
  pages={245411},
  year={2023},
  publisher={APS}
}

@article{chiritescu2007ultralow,
  title={Ultralow thermal conductivity in disordered, layered WSe2 crystals},
  author={Chiritescu, Catalin and Cahill, David G and Nguyen, Ngoc and Johnson, David and Bodapati, Arun and Keblinski, Pawel and Zschack, Paul},
  journal={Science},
  volume={315},
  number={5810},
  pages={351--353},
  year={2007},
  publisher={American Association for the Advancement of Science}
}

@article{li2022layered,
  title={Layered thermoelectric materials: Structure, bonding, and performance mechanisms},
  author={Li, Zhou and Xiao, Chong and Xie, Yi},
  journal={Applied Physics Reviews},
  volume={9},
  number={1},
  year={2022},
  publisher={AIP Publishing}
}

@article{wang2017thermal,
  title={Thermal properties of two dimensional layered materials},
  author={Wang, Yuxi and Xu, Ning and Li, Deyu and Zhu, Jia},
  journal={Advanced Functional Materials},
  volume={27},
  number={19},
  pages={1604134},
  year={2017},
  publisher={Wiley Online Library}
}

@book{srivastava2022physics,
  title={The physics of phonons},
  author={Srivastava, Gyaneshwar P},
  year={2022},
  publisher={CRC press}
}

@article{tantardini2021thermochemical,
  title={Thermochemical electronegativities of the elements},
  author={Tantardini, Christian and Oganov, Artem R},
  journal={Nature communications},
  volume={12},
  number={1},
  pages={2087},
  year={2021},
  publisher={Nature Publishing Group UK London}
}

@article{nissimagoudar2020lattice,
  title={Lattice thermal transport in monolayer group 13 monochalcogenides MX (M= Ga, In; X= S, Se, Te): Interplay of atomic mass, harmonicity, and lone-pair-induced anharmonicity},
  author={Nissimagoudar, Arun S and Rashid, Zahid and Ma, Jinlong and Li, Wu},
  journal={Inorganic Chemistry},
  volume={59},
  number={20},
  pages={14899--14909},
  year={2020},
  publisher={ACS Publications}
}

@article{long2025theoretical,
  title={Theoretical study of electronic structure, lone pair localization, and electronic transport properties of unconventional bulk and 2D $\gamma$-SnSe and $\gamma$-SnS},
  author={Long, Nguyen Truong and Huy, Huynh Anh and Mishra, Neeraj and Makov, Guy},
  journal={RSC advances},
  volume={15},
  number={21},
  pages={16358--16374},
  year={2025},
  publisher={Royal Society of Chemistry}
}

@article{zhang2026low,
  title={Low lattice thermal conductivity in thermoelectric PbBi 2 Te 4 induced by double lone pair electrons},
  author={Zhang, Jingyi and Bai, Shulin and Sun, Shuai and Zhang, Pengfei and Ai, Peng and Peng, Junhao and Liang, Yanwei and Tang, Shuwei and Dong, Huafeng},
  journal={Physical Review B},
  volume={113},
  number={16},
  pages={165203},
  year={2026},
  publisher={APS}
}

@article{parashchuk2022ultralow,
  title={Ultralow lattice thermal conductivity and improved thermoelectric performance in Cl-doped Bi2Te3--x Se x alloys},
  author={Parashchuk, Taras and Knura, Rafal and Cherniushok, Oleksandr and Wojciechowski, Krzysztof T},
  journal={ACS applied materials \& interfaces},
  volume={14},
  number={29},
  pages={33567--33579},
  year={2022},
  publisher={ACS Publications}
}

@article{wu2026lone,
  title={From lone-pair electrons to dual phonon channels: Unraveling Te-dominated transport in monolayer Sb 2 Te 3},
  author={Wu, Kai and Zhou, Ran and Shi, Hongliang and Duan, Yifeng},
  journal={New Journal of Physics},
  year={2026}
}

@article{adhidewata2022thermoelectric,
  title={Thermoelectric properties of semiconducting materials with parabolic and pudding-mold band structures},
  author={Adhidewata, Jyesta M and Nugraha, Ahmad RT and Hasdeo, Eddwi H and Estell{\'e}, Patrice and Gunara, Bobby E},
  journal={Materials Today Communications},
  volume={31},
  pages={103737},
  year={2022},
  publisher={Elsevier}
}

@article{ganose2021ifermi,
  title={IFermi: A python library for Fermi surface generation and analysis},
  author={Ganose, Alex M and Searle, Amy and Jain, Anubhav and Griffin, Sin{\'e}ad M},
  journal={Journal of Open Source Software},
  volume={6},
  number={59},
  pages={3089},
  year={2021}
}

@article{ryu2021thermoelectric,
  title={Thermoelectric degrees of freedom determining thermoelectric efficiency},
  author={Ryu, Byungki and Chung, Jaywan and Park, SuDong},
  journal={Iscience},
  volume={24},
  number={9},
  year={2021},
  publisher={Elsevier}
}

@article{ryu2025thermoelectric,
  title={Thermoelectric algebra made simple for thermoelectric generator module performance prediction under constant Seebeck coefficient approximation},
  author={Ryu, Byungki and Chung, Jaywan and Park, SuDong},
  journal={Journal of Applied Physics},
  volume={137},
  number={5},
  year={2025},
  publisher={AIP Publishing}
}

@article{zhao2020chemical,
  title={Chemical doping of organic semiconductors for thermoelectric applications},
  author={Zhao, Wenrui and Ding, Jiamin and Zou, Ye and Di, Chong-an and Zhu, Daoben},
  journal={Chemical Society Reviews},
  volume={49},
  number={20},
  pages={7210--7228},
  year={2020},
  publisher={Royal Society of Chemistry}
}

@article{ayachi2024high,
  title={High-Performance thermoelectric devices made faster: interface design from first principles calculations},
  author={Ayachi, Sahar and Park, Sungjin and Ryu, Byungki and Park, SuDong and Mueller, Eckhard and de Boor, Johannes},
  journal={Advanced Physics Research},
  volume={3},
  number={1},
  pages={2300077},
  year={2024},
  publisher={Wiley Online Library}
}

@article{lee2019fine,
  title={Fine tuning of Fermi level by charged impurity-defect cluster formation and thermoelectric properties in n-type PbTe-based compounds},
  author={Lee, Min Ho and Park, Sungjin and Lee, Jae Ki and Chung, Jaywan and Ryu, Byungki and Park, Su-Dong and Rhyee, Jong-Soo},
  journal={Journal of Materials Chemistry A},
  volume={7},
  number={27},
  pages={16488--16500},
  year={2019},
  publisher={Royal Society of Chemistry}
}

@article{phono3py,
  title = {Distributions of phonon lifetimes in Brillouin zones},
  author = {Togo, Atsushi and Chaput, Laurent and Tanaka, Isao},
  journal = {Phys. Rev. B},
  volume = {91},
  issue = {9},
  pages = {094306},
  numpages = {31},
  year = {2015},
  month = {Mar},
  publisher = {American Physical Society},
  doi = {10.1103/PhysRevB.91.094306},
}

@article{phonopy-phono3py-JPCM,
  author  = {Togo, Atsushi and Chaput, Laurent and Tadano, Terumasa and Tanaka, Isao},
  title   = {Implementation strategies in phonopy and phono3py},
  journal = {J. Phys. Condens. Matter},
  volume  = {35},
  number  = {35},
  pages   = {353001},
  year    = {2023},
  doi     = {10.1088/1361-648X/acd831}
}

@article{PhysRevLett.110.265506,
  title = {Direct Solution to the Linearized Phonon Boltzmann Equation},
  author = {Chaput, Laurent},
  journal = {Phys. Rev. Lett.},
  volume = {110},
  issue = {26},
  pages = {265506},
  numpages = {5},
  year = {2013},
  month = {Jun},
  publisher = {American Physical Society},
  doi = {10.1103/PhysRevLett.110.265506},
  url = {https://link.aps.org/doi/10.1103/PhysRevLett.110.265506}
}

\end{document}